\documentclass[pra,aps,10pt,twocolumn,amsmath,amssymb,floatfix,footinbib,bibnotes,longbibliography,usenames,dvipsnames]{revtex4-1}
\pdfoutput=1

\usepackage[colorlinks=true,citecolor=blue,linkcolor=magenta]{hyperref}
\usepackage{amsmath}
\usepackage{graphicx}
\usepackage{changepage}
\usepackage{amsthm, amssymb}
\usepackage{floatflt}
\usepackage{float}
\usepackage{array}
\usepackage[usenames,dvipsnames]{color}
\usepackage{mathtools}
\usepackage[normalem]{ulem}

\def \w{\omega}

\def \t{\tau}
\def \beq{\begin{equation}}
\def \beqn{\begin{align}}
\def \eeq{\end{equation}}
\def \eeqn{\begin{align}}

\def  \G{\Gamma}
\def  \q0{\frac{\w_k^2}{\G}+z}
\def  \qe{q_{\text{EA}}}
\def  \qd{Q_d}

\def  \q0{\frac{\w_k^2}{\G}+z}

\newcommand{\ginv}{\big(\tilde{Q}^{-1}\big)_{ab}}
\newcommand{\qr}{Q_{\text{reg}}} 

\newcommand{\qt}{\tilde{q}_{{\rm EA}}}
\newcommand{\ci}{{\rm i}}

\def \titlename {Sachdev-Ye-Kitaev Model in  a Quantum Glassy Landscape}

\def \authornames{Surajit Bera$^{1}$, Jorge Kurchan$^2$, and  Marco Schir\`o$^1$}
\def \affiliations{$^1$JEIP, UAR 3573 CNRS, Collège de France, PSL Research University,\\ 11 Place Marcelin Berthelot, 75321 Paris Cedex 05, France\\$^{2}$Laboratoire de Physique de l’École Normale Supérieure, ENS, Université PSL, CNRS, Sorbonne Université, Université de Paris, F-75005 Paris, France
}

\begin{document}
	\title{\titlename}
\author{\authornames}
	\affiliation{\affiliations}
	\date\today

\begin{abstract}
We study a generalization of `Yukawa models'  in which Majorana fermions, interacting via all-to-all random couplings as in the Sachdev-Ye-Kitaev (SYK) model, are parametrically coupled to disordered bosonic degrees of freedom described by a quantum $p-$spin model. The latter has its own non-trivial dynamics leading to quantum paramagnetic (or liquid) and glassy phases.
At low temperatures this setup results in SYK behavior  within each  metastable state of a rugged bosonic free energy landscape, the effective fermionic couplings being different for each metastable state. 
We show that the boson-fermion coupling enhances the stability of the quantum spin-glass phase and strongly modifies the imaginary-time Green's functions of both sets of degrees of freedom. In particular,
in the quantum spin glass phase, the imaginary-time dynamics is turned from a fast exponential decay characteristic of a gapped phase into a much slower dynamics. In the quantum paramagnetic phase, on the other hand, the fermions imaginary-time dynamics get strongly modified and the critical SYK behavior is washed away.
\end{abstract}

\maketitle 
\section{Introduction}

In recent years, the Sachdev-Ye-Kitaev (SYK) model has attracted significant attention in a wide range of research fields~\cite{kitaev2015quantum_holography, kitaev2015quantum_holography_part2, PhysRevD.94.106002, chowdhury2017onset, gu2017energy, murugan2017susysyk, Hosur2016}. The SYK model features random all-to-all interactions among $q$  Majorana or complex fermions.  SYK and its variants exhibit several remarkable features: they provide a solvable example of non-Fermi liquid behavior~\cite{RevModPhys.94.035004, PhysRevResearch.4.013145}, offer insight into the mechanism behind strange and Planckian metallicity~\cite{PhysRevLett.123.066601, altman2019strangemetal, patel2022planckian, PhysRevLett.119.216601, RevModPhys.94.041002}, saturate the bound on quantum chaos~\cite{chen2017sykchaos, EhudSumilan, chowdhury2017onset}, and capture essential aspects of black hole dynamics through holographic correspondence~\cite{Kitaev2018, Lau_2024}.
Its solvability in the large-$N$ limit, where $N$ denotes the number of fermionic flavors, thus provides a powerful framework for exploring these diverse physical phenomena, in fields ranging from strongly correlated electron systems in condensed matter to holographic dualities in quantum gravity.

The SYK model originates as an effective theory from a spin-glass model,  the quantum random Heisenberg model~\cite{SachdevYe1993, GeorgesParcolletSachdev2001, PhysRevB.59.5341}. It can also be viewed as a fermionic analogue of another paradigmatic and exactly solvable mean-field spin-glass model—the quantum $p$-spin glass model~\cite{KirkpatrickThirumalai, Crisanti1992, Nieuwenhuizen1995, LeticiaGustavo, Leticia2001, Dobrosavljevic_1990}. This model involves random all-to-all couplings among $p$ bosonic degrees of freedom, shares similar large-$N$ saddle-point equations, and exhibits a form of time-reparametrization symmetry~\cite{sachdev2024glass, FacoettiJorge, QuantumGlass}, similar to the features of the SYK model. A key characteristic of such mean-field spin-glass systems is their rugged potential energy landscape, with the number of metastable minima or maxima scaling exponentially with system size $N$. The bosonic degrees of freedom can remain trapped in these metastable states for  divergent times in the thermodynamic limit, effectively breaking ergodicity below a critical temperature and leading to spin-glass order.

\begin{figure}[htb]
    \centering
    \includegraphics[width=0.9\linewidth]{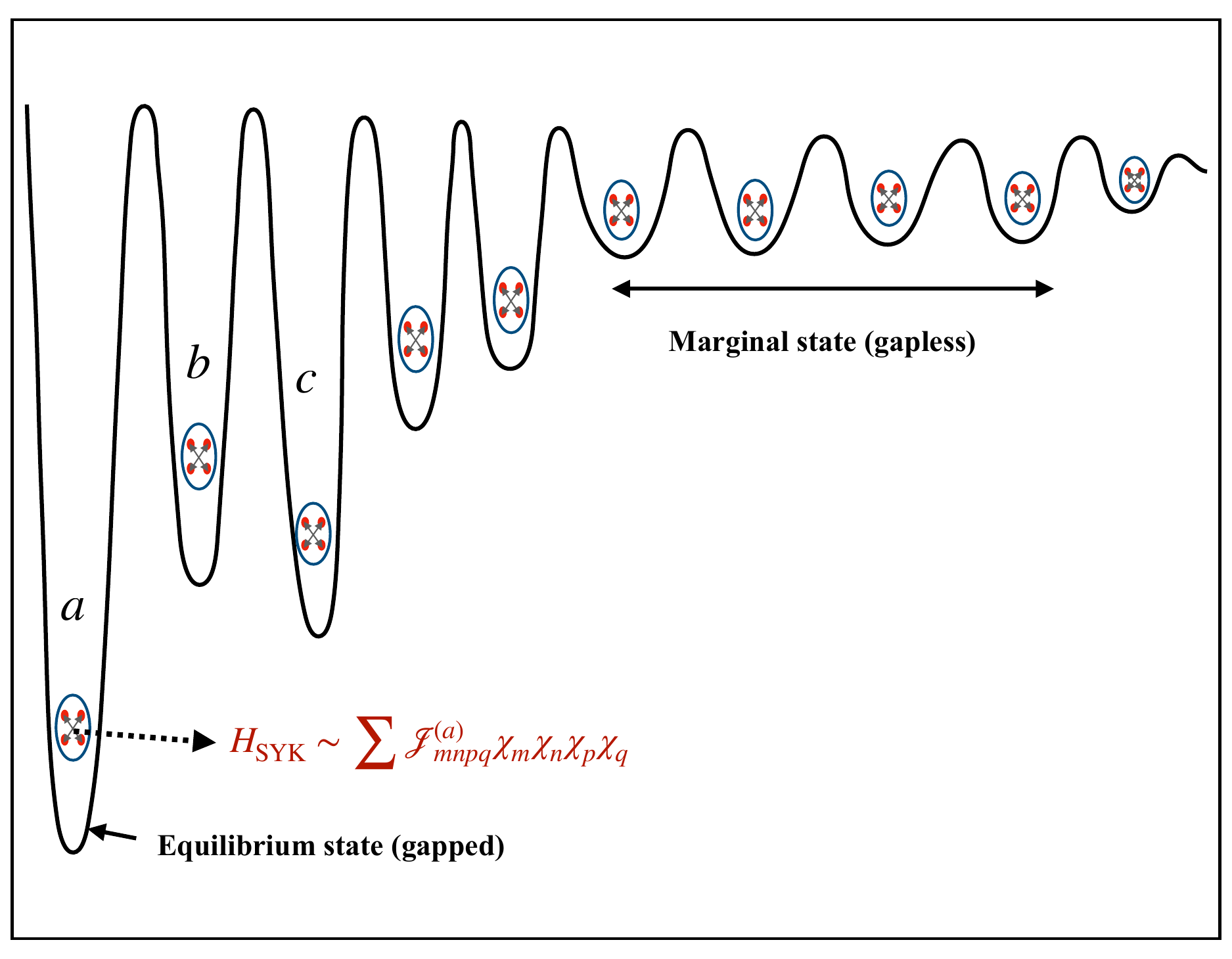}
    \caption{\textbf{SYK Dots in the Energy Landscape of the Spin-Glass Model:}  
A schematic cartoon of the typical energy landscape of the $p$-spin-glass model is shown.  
The complex free-energy landscape consists of equilibrium states (gapped), an exponential number of metastable states, and marginal (threshold) states that are gapless.  
Each SYK dot is located in a different valley of this landscape, with a distinct fluctuating glassy background determined by the specific state $a, b, c, \ldots$.  
At very low $T$, the SYK dots are expected to occupy the low-lying minima (equilibrium states) in the semi-classical limit of the $p$-spin-glass model.
}
    \label{fig:landscape}
\end{figure}

The p-spin model has been extensively studied as a solvable glass \cite{Kirkpatrick1987PRB,KirkpatrickThirumalai,Wolynes1997JRNIST,Xia2000PNAS}. 
Its quantum version  itself exhibits a rich phase diagram \cite{Leticia2001} (see also \cite{goldschmidt1990solvable,giamarchi1996variational}). At low temperatures, the bosonic degrees of freedom condense into a spin-glass phase with broken ergodicity, characterized by one-step replica symmetry breaking (1-RSB) in the replica formalism framework. In contrast, at higher temperatures or under strong quantum fluctuations, the system enters a quantum paramagnetic phase with replica-diagonal order parameters. 
The phase diagram in the $\Gamma$–$T$ plane -- where $\Gamma$ quantifies the strength of quantum fluctuations and $T$ is the temperature of the bosonic system -- has been  studied in literature; we refer the reader to Ref.~\cite{Leticia2001} for a detailed discussion. The low-temperature metastable states are organized as in Fig \ref{fig:landscape}: there are the very lowest in free energy that dominate the Gibbs-Boltzmann measure (and are described by a standard replica calculation), plus an exponentially increasing number with higher and higher free energies, up to a `threshold' level where states are marginal, i.e. gapless: these states dominate the out of equilibrium dynamics after a quench \cite{cugliandolo1993analytical,LeticiaGustavo}. To avoid confusion, throughout this paper we shall use the word `state $a$' to denote a distribution $\rho_a$ that is stationary under dynamics and not decomposable into smaller  stable ones (an object that could as well be classical), while one shall use `eigenstate' to denote  a quantum eigenstate of a Hamiltonian.
The model has two types of paramagnetic solutions, namely a `classical' paramagnet exits in high temperature or small $\Gamma$ limit, and  a `quantum' paramagnet exists in low temperature at very high value of $\Gamma$.

In this work, we investigate how Majorana fermions with SYK-type all-to-all interactions evolve under the influence of background fluctuations arising from a disordered bosonic environment described by the quantum spherical $p$-spin glass model. To this end, we study a model of disordered fermion-boson interactions involving random all-to-all couplings. We consider a fermion-boson interaction term in which Majorana fermions interact via SYK-type four-fermion interactions that are parametrically modulated by their coupling to bosonic degrees of freedom governed by a quantum spherical $p$-spin glass model. The quantum fluctuations of bosonic degrees of freedom directly influence the effective couplings among fermions. This setup captures the intuitive picture of SYK fermions evolving within the complex landscape of a mean-field spin-glass potential, effectively confined in different metastable valleys corresponding to local minima of the bosonic free energy (see Figure \ref{fig:landscape} ). At very low temperatures, the fermions become trapped in these minima and experience background fluctuations controlled by the spin-glass order.

Our  aim is to address the following questions: How do fermions evolve across different regimes of the spin-glass phase diagram? How does fermionic feedback reshape the bosonic order parameters and correlations? Our focus and motivations differ from those in earlier studies of SYK coupled with bosonic fields. Prior work has studied Yukawa-type interactions\cite{PhysRevLett.124.017002}, where fermions interact with critical bosonic fields associated with paramagnetic order parameters. Such models have been shown to give rise to a variety of strongly correlated phases, including superconductivity~\cite{ClassenAndrew, PhysRevB.100.115132, ValentinisSchmalian}, non-Fermi liquid behavior with critical Fermi surfaces~\cite{IlyaSubir2021, Yuxuan2020}, and quantum criticality~\cite{Yuxuan2020, YuxuanAndrew}. In parallel, supersymmetric extensions of the SYK model have also been investigated, where SYK fermions couple to their supersymmetric bosonic partners~\cite{benini2024mathcaln2sykmodelsdynamical, murugan2017susysyk, FuMaldacenaSubir, Biggs_2024}. In contrast, here we study the coupling of fermions to both non-critical (equilibrium glass) and critical (marginal glass\cite{MarkusMuller}), glassy bosonic fields, revealing all to all coupled fermionic dynamics under fluctuating background, and uncovering novel feedback effects of fermion to spin-glass order.


The boson-fermion coupling has two main effects on the physics of our model. First, it modifies the phase diagram by enhancing the stability of the thermodynamic quantum spin-glass phase at finite quantum fluctuation $\Gamma$. Second, it strongly affects the imaginary-time correlation functions of both sets of degrees of freedom. In particular, in the quantum spin-glass phase the bosonic correlator at low-temperature is strongly renormalised and its exponential decay in imaginary-time washed out. In the quantum paramagnet phase on the other hand is the fermionic sector dynamics which is mostly affected and the SYK critical behavior is turned into a slow dynamics characterised by a flat correlation function.

The manuscript is organized as follows. In Sec.~\ref{sec:Model}, we introduce the coupled fermion-boson model, derive the large-$N$ saddle-point equations, and briefly review key features of the decoupled spherical spin-glass and SYK models. In Sec.~\ref{sec:spinglass}, we present results for the spin-glass phase of the coupled system, followed by an analysis of the paramagnetic phase in Sec.~\ref{sec:PMphase}. We conclude with a discussion in Sec.~\ref{sec:discussion}. Additional technical details are provided in the Appendix.

\section{Model}\label{sec:Model}
We consider the following Hamiltonian:
\begin{align}\label{eqn:full-hamiltonian}
    \mathcal{H} &= \mathcal{H}_{\rm b}[s] + \mathcal{H}_{\rm int}[s,\chi],
\end{align}
where $\mathcal{H}_{\rm int}[s,\chi]$ denotes the interaction term between bosons ($s$) and Majorana fermions ($\chi$). The bosonic part $\mathcal{H}_{\rm b}$ is given by the Hamiltonian of a quantum spherical $p$-spin glass model:
\begin{align}
    \mathcal{H}_{\rm b} &= \sum_{i} \frac{\pi_i^2}{2M} + \sum_{i<j<k} J_{ijk} s_i s_j s_k.
\end{align}
Here, $s_i$ and $\pi_i$ are canonically conjugate variables satisfying the commutation relation $[s_i, \pi_j] = {\rm i} \hbar \delta_{ij}$, where $\hbar$ is the Planck constant. The coordinate $s_i$ is continuous, taking values over the entire real line, $(-\infty, +\infty)$. In this work, we focus on the case $p = 3$, i.e., interactions involving three bosons. The all-to-all random couplings $J_{ijk}$ are real variables drawn from a Gaussian distribution with zero mean and variance
\[
\sigma^2_J = \langle J^2_{ijk} \rangle = \frac{J^2 \cdot 3!}{2 N^{2}}.
\]
{The 
$1/N^2$
 scaling of the coupling variance ensures extensivity of the free energy, the correct dynamic evolution.} To ensure the model remains non-trivial, we impose a spherical constraint:
\[
\sum_i s_i^2 = N,
\]
where $N$ is the total number of bosonic modes. In the context of the $p$-spin glass model, the degrees of freedom $s_i$ are often referred to as spin variables. Throughout this work, we use the terms ``spin" and ``boson" interchangeably to refer to the coordinate $s_i$.

The interaction between the bosonic coordinates $s_i$ and the Majorana fermions $\chi_i$ is given by:
\begin{align}\label{eq:Hint0}
    \mathcal{H}_{\rm int} &= \sum_{i_1,...,i_r} \sum_{mnpq} V^{mnpq}_{i_1...i_r} \chi_m \chi_n \chi_p \chi_q\, s_{i_1} ... s_{i_r} 
\end{align}
We do not expect the order $r>0$ of the bosonic part to have a great impact, since the glassiness of bosons is given by the pure boson term. Here we concentrate on $r=4$: 
\begin{align}\label{eq:Hint}
    \mathcal{H}_{\rm int} &= \sum_{ijkl} \sum_{mnpq} V^{mnpq}_{ijkl} \chi_m \chi_n \chi_p \chi_q\, s_i s_j s_k s_l.
\end{align}
The interaction tensor $V^{mnpq}_{ijkl}$ is a real variable drawn from a Gaussian distribution with zero mean. It is symmetric under the exchange of any pair of bosonic indices $(i, j, k, l)$ and antisymmetric under the exchange of any pair of fermionic indices $(m, n, p, q)$. We assume that the number of bosonic and fermionic degrees of freedom is equal, i.e., both sets of indices range from $1$ to $N$. The variance of $V^{mnpq}_{ijkl}$ is given by:
\begin{align}
    \sigma^2_V = \langle (V^{mnpq}_{ijkl})^2 \rangle = \frac{4! \cdot 3! \cdot V^2}{2 N^7}.
\end{align}
{It is chosen to ensure a well-defined large-$N$ limit and an extensive free energy, $\mathcal{O}(N)$. The scaling of $\sigma_V$ can be fixed by requiring the disorder-averaged interaction energy fluctuations to be extensive, namely
\begin{equation}
\overline{\mathrm{Tr}\big(\mathcal{H}_{\rm int}^2\big)} \sim \mathcal{O}(N).
\end{equation}
Since different disorder realizations are statistically independent,
\begin{equation}
\overline{V^{mnpq}_{ijkl} V^{m'n'p'q'}_{i'j'k'l'}} \propto
\delta_{ii'}\cdots\delta_{ll'}\,\delta_{mm'}\cdots\delta_{qq'},
\end{equation}
the disorder average of $\mathrm{Tr}(\mathcal{H}_{\rm int}^2)$ receives contributions only from identical index contractions. Consequently, the sum over the eight free indices $i,j,k,l,m,n,p,q$ yields a factor of $N^8$, leading to
\begin{equation}
\overline{\mathrm{Tr}\big(\mathcal{H}_{\rm int}^2\big)} \sim \sigma_V^2\, N^8.
\end{equation}
Requiring this quantity to scale as $\mathcal{O}(N)$ fixes the variance to scale as $\sigma_V^2 \sim N^{-7}$. The remaining numerical prefactors are conventional combinatorial factors. With these choices, both the energy fluctuations and the free energy are extensive, while the corresponding self-energies remain $\mathcal{O}(1)$ in the large-$N$ limit.}

The interaction term in Eq.~\eqref{eq:Hint}, involving four Majorana fermions coupled to bosonic degrees of freedom, plays a crucial role in mimicking SYK-like fermionic dynamics in the presence of a fluctuating,  bosonic background in its glassy phase. In particular, it leads, in the classical limit of the quantum spin-glass model, to an effective fermionic Hamiltonian of the form:
\begin{align}\label{eq:SYKHam}
    H_{\rm fermion} \sim \mathcal{H}_{\rm SYK} = \sum_{mnpq} \mathcal{J}_{mnpq} \chi_m \chi_n \chi_p \chi_q.
\end{align}
where the couplings $\mathcal{J}_{mnpq}$ are dynamically modulated by the fluctuating spin configuration $\vec{s}$ as:
\begin{align}
    \mathcal{J}_{mnpq}(t) = \sum_{ijkl} V^{mnpq}_{ijkl} s_i s_j s_k s_l.
\end{align}

In the original SYK model,  the couplings are static and are drawn from a Gaussian distribution with zero mean and variance $\langle (\mathcal{J}_{mnpq})^2 \rangle = \mathcal{J}^2 / 2N^3$. To the extent that the  $s_i$ may be approximated as  taking  constant, static values, we are essentially back to the original SYK, with modified interactions, {\it but these are different in each  bosonic valley}. In any case, this must be  interpreted with care in the quantum regime, since the bosonic variables $s_i(\tau)$ are quantum fields that fluctuate in imaginary time $\tau$, even at zero temperature. In a later section, we will discuss how these quantum spin fluctuations modify the fermionic saddle-point equations compared to those of the original SYK model, and under what conditions the standard SYK saddle point for fermions can be recovered.

\subsection{Large-$N$ saddle point equations}
The thermodynamics of the system described by the above Hamiltonian can be derived from the corresponding free energy functional. For disordered systems of this type, the free energy is self-averaging, meaning it becomes independent of a specific realization of disorder in the thermodynamic limit. Consequently, the disorder-averaged free energy at temperature $T$ is given by:
\begin{align}
    F &= -T\, \overline{\log Z},
\end{align}
where the overline $(\overline{\cdots})$ denotes the average in different realizations of the disorder variables $J_{ijk}$ and $V^{mnpq}_{ijkl}$. 

To evaluate the disorder-averaged logarithm of the partition function, we employ the replica trick, write down the partition function for $r$ copies (replicas) of the Hamiltonian Eq. \ref{eqn:full-hamiltonian} in path integral language, and then perform the disorder average as described in detail in Appendix~\ref{app:derivation_saddle}. The effective partition function for $r$-replica reads as follows:
\begin{align}
    \overline{Z^r} &= \int \mathcal{D}[G, \Sigma, Q, \Pi] e^{-S_{\rm eff}[G, \Sigma, Q, \Pi]}
\end{align}
where the effective action is given by: 
\onecolumngrid
\begin{align}
S_{\rm eff} &= -N \operatorname{Tr} 
             \ln \big( -G^{-1}_{0,ab} + \Sigma_{ab} \big) 
             - \frac{N}{2} \operatorname{Tr} \ln \big( Q^{-1}_{0,ab} -\Pi_{ab} \big) 
             + \frac{NV^2}{4}\int {d\tau d\tau'}\sum_{ab} G^{4}_{ab}(\tau,\tau') Q^{4}_{ab}(\tau,\tau') \notag\\ & + \frac{N J^2}{4}\int {d\tau d\tau'}\sum_{ab}  Q^{ 3}_{ab}(\tau,\tau') 
            -N \int {d\tau d\tau'} \sum_{ab}\Sigma_{ab}(\tau,\tau') G_{ab}(\tau,\tau')
             - \frac{N}{2} \int {d\tau d\tau'} \sum_{ab} \Pi_{ab}(\tau,\tau') Q_{ab}(\tau,\tau')             
\end{align}
\twocolumngrid
Here, $\text{Tr}$ is trace over both replica indicse and $\tau$. 
In the above, replica indices $a, b$ runs from 1 to $r$, and at the end of calculation, one need to take $r\to 0$ limit carefully. {In the above, $G^{-1}_{0,ab}(\tau) = -\delta(\tau) \delta_{ab}\partial_{\tau} $ denotes the inverse free fermionic propagator, while 
$
Q^{-1}_{0,ab}(\tau) = \delta(\tau)\delta_{ab}\left(-M \partial^2_{\tau} + z \right)
$
is the inverse free bosonic propagator. Here, $z$ is the Lagrange multiplier that enforces the spherical constraint in the path-integral formulation.} 

The functions $Q_{ab}(\tau)$ and $G_{ab}(\tau)$ are the order parameter fields or Green's functions for the bosonic and fermionic degrees of freedom, respectively, defined as:
\begin{align}
    G_{ab}(\tau) &= \frac{1}{N} \sum_i \langle \chi_{ia}(\tau) \chi_{ib}(0) \rangle, \\
    Q_{ab}(\tau) &= \frac{1}{N} \sum_i \langle s_{ia}(\tau) s_{ib}(0) \rangle.
\end{align}
 The quantities $\Pi_{ab}(\tau)$ and $\Sigma_{ab}(\tau)$  are the bosonic and fermionic self-energies in imaginary time, conjugate fields to $Q_{ab}(\tau)$ and $G_{ab}(\tau)$, respectively. The Matsubara transforms of $Q_{ab}(\tau)$ are defined as:
\begin{align}
    \tilde{Q}_{ab}(\omega_k) &= \int_0^{\beta} d\tau\, e^{i\omega_k \tau} Q_{ab}(\tau), \\
    Q_{ab}(\tau) &= T \sum_{\omega_k} e^{-i\omega_k \tau} \tilde{Q}_{ab}(\omega_k),
\end{align}
and similar expressions hold for $\tilde{G}_{ab}(\omega_n)$ and the self-energy functions. In the above, $\beta = 1/T$ is the inverse temperature. From here on, we measure the imaginary time $\tau$ in units of $1/J$ (setting $\hbar = 1$). 

In the limit of large-$N$, the set of saddle point equations in matsubara frequency space is given as follows:
\begin{align}
    \tilde{Q}^{-1}_{ab}(\omega_k) &= (\omega_k^2/\Gamma + z)\, \delta_{ab} - \tilde{\Pi}_{ab}(\omega_k),\label{eq:Qabwn} \\
    \Pi_{ab}(\tau) &= \frac{3J^2}{2} Q_{ab}^2(\tau) + V^2 Q_{ab}^3(\tau) G_{ab}^4(\tau), \label{eq:Piabwn} \\
    \tilde{G}^{-1}_{ab}(\omega_n) &= -i\omega_n \delta_{ab} - \tilde{\Sigma}_{ab}(\omega_n), \label{eq:Gabwn}\\
    \Sigma_{ab}(\tau) &= V^2 Q_{ab}^4(\tau) G_{ab}^3(\tau).\label{eq:sigmaabwn}
\end{align}
Here, $\omega_k = 2k \pi  T $ and $\omega_n = (2n+1)\pi  T$ (setting $k_B=1$) are bosonic and fermionic Matsubara frequencies, respectively, with $n, k \in \mathbb{Z}$ and $k_B$ the Boltzmann constant. The parameter $\Gamma=1/MJ$ controls the quantum fluctuation. In the following, we discuss the structure of the fermionic and bosonic order parameter matrix.

\subsection{Structure of order parameter for bosons and fermions}

\begin{figure}
    \centering
    \includegraphics[width=1\linewidth]{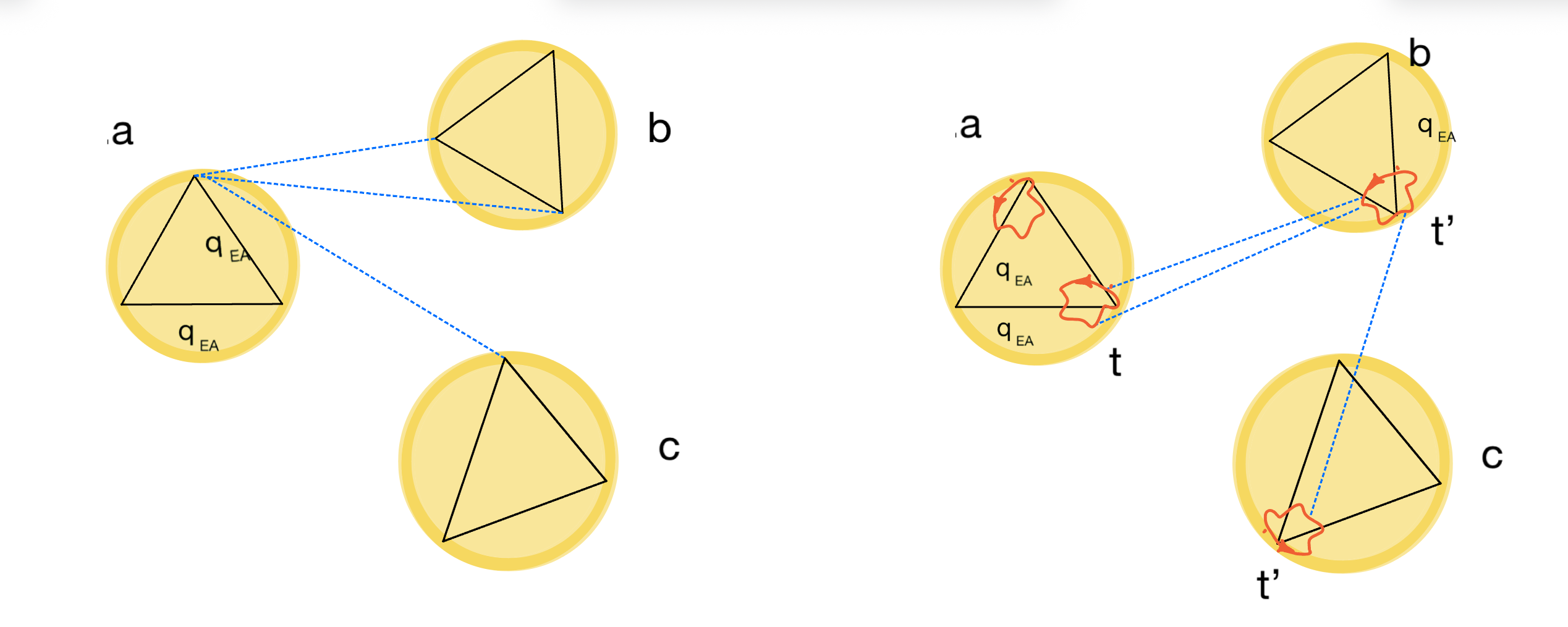}
    \caption{Hierarchical (Parisi) structure of configurations. Left: classical. Within a state {\bf a,b,c ...} almost all configurations are at the same distance from one another $q_{EA}$, a known property of high-dimensional sets. Similarly, the mutual distance between any two configurations in different states is also the same, but different from $q_{EA}$ . Right: quantum. The situation is the same, but now we have a `collar' of configurations, the imaginary-time trajectory, whose `beads' are labeled by $t$. The organization now extends one step inwards: two different times of a trajectory are at the same distance from any time of a trajectory in another state.  
    }
    \label{states}
\end{figure}

The Parisi ansatz for the order parameters of a quantum problem $Q_{ab}(\tau)$, $G_{ab}(\tau)$, where $\tau$ is the Matsubara time  is shown in figure \ref{states}. As in any quantum problem, we may think of trajectories  as a closed `collar' with `beads' of indices $\tau$, living {\em within a  state} in the energy landscape.  A natural generalization of ultrametricity is that two  ``beads" from a given chain are at the same distance of each monomer of a chain in another state see Fig. \ref{states}.
This is the same as saying  that only diagonal terms $Q_{aa}$ and $G_{aa}$ may depend on $\tau$. In particular, this means that $G_{ab}$ necessarily is zero for $a\neq b$, because it concerns different fermions at equal times. 

In what follows, we will consider both the replica-diagonal ansatz (i.e., $Q_{ab}=0$ for $a \neq b$) and the one-step replica symmetry breaking (1-RSB) ansatz for $Q_{ab}$. The fermionic Green’s function $G_{ab}(\tau)$ will be taken in a replica-diagonal form throughout, as explained above. 
Before analyzing the results for the coupled system, we first review the phase structure of the decoupled bosonic model ($V=0$). We then discuss how the fermion--boson interaction modifies the structure of the saddle-point equations in the coupled case. Finally, we compare the saddle-point equations of the pure SYK model with those of the all-to-all coupled fermions in the presence of bosonic interaction.


\subsubsection{Phase structure of decoupled boson}

In the decoupled case, i.e., $V = 0$, the spherical quantum $p$-spin glass model (with $p = 3$) is well studied and is known to host a rich phase diagram~\cite{Leticia2001}. The replica-diagonal $Q_{aa}$ paramagnetic phase exists in the region of relatively high temperature and large quantum fluctuations $\Gamma$. 
Interestingly, the $p = 3$ model exhibits a particular temperature $T_p \approx 0.167$, below which two  distinct paramagnetic solutions coexist within a finite range of the quantum parameter, $\Gamma_{c1} < \Gamma < \Gamma_{c2}$. They are now everywhere continuously connected, just like a liquid-gas transition. They prolong into two regimes: one at large quantum fluctuations ($\Gamma \gg 1$), referred to as the \emph{quantum paramagnet}, and the other at small quantum fluctuations ($\Gamma \ll 1$), referred to as the \emph{classical paramagnet}. For temperatures $T > T_p$, only a single paramagnetic solution exists for each value of $\Gamma$, and the classical and quantum paramagnetic regimes are smoothly connected.

At low temperatures, a one-step replica symmetry broken (1-RSB) spin-glass phase emerges. The bosonic order parameter in the 1-RSB phase takes the following form:
\begin{align}\label{eq:1RSBQab}
    Q_{ab}(\tau) &= \left(Q_{aa}(\tau) - q_{\rm EA}\right) \delta_{ab} + q_{\rm EA} \, \epsilon_{ab},
\end{align}
where $\epsilon_{ab} = 1$ if replicas $a$ and $b$ belong to the same diagonal block, and $\epsilon_{ab} = 0$ otherwise. The parameter $q_{\rm EA}$ represents the off-diagonal component within the replica blocks and is known as the Edwards–Anderson (EA) order parameter. Therefore, the order parameter matrix $Q_{ab}$ $(n\times n)$ has a   block diagonal structure with each block has dimension $m\times m$ $(m\leq n)$.  Depending on how the replica symmetry breaking breakpoint parameter $m$ is determined, two types of spin-glass phases can be distinguished:    (a) \textit{Thermodynamic spin glass:} The breakpoint $m$ is determined by minimizing the free energy, ensuring thermodynamic stability. 
(b) \textit{Marginal spin glass:} The breakpoint $m$ is fixed by requiring that the replicon eigenvalue vanishes, leading to a marginally stable gapless solution. (This alternative has been advocated in Ref. \cite{giamarchi1996variational} for quantum models, and has become widespread, but it has to be remembered that it does not correspond to thermal equilibrium). Referring to Figure \ref{fig:landscape}, tuning the value of $m$ between these points we capture metastable states ranging from the lowest, that dominate the thermodynamic measure, to the highest `threshold states', which are gapless and are the ones that are relevant for most of the out of equilibrium situations.

\subsubsection{{Saddle point  structure for coupled boson} }
{

As in the decoupled $(V=0)$ spin-glass phase, we assume a one-step replica symmetry breaking (1RSB) order parameter for the coupled model. The 1RSB order parameter can be conveniently decomposed into a regular part,
\begin{equation}
Q_{\rm reg}(\tau)=Q_{d}(\tau)-q_{\rm EA},
\end{equation}
and an off-diagonal Edwards--Anderson component $q_{\rm EA}$. In addition, the breakpoint parameter $m$ must be specified to fully characterize the 1RSB solution.

As discussed in the decoupled case, we consider two types of spin-glass phases: (a) the thermodynamic spin-glass phase and (b) the marginal spin-glass phase. In the following, we examine how the resulting set of equations for $Q_{\rm reg}(\omega_k)$, $q_{\rm EA}$, and $m$ is modified by the fermion--boson coupling, and how it compares with the corresponding equations in the decoupled model. As we show below, the equation for $Q_{\rm reg}(\omega_k)$ acquires a fermionic feedback through the fermion--boson interaction term. However, since the fermion--boson coupling is replica diagonal, it does not modify the saddle-point equation for $q_{\rm EA}$ nor the set of equations that determine the breakpoint parameter $m$ in either the thermodynamic or marginal spin-glass phase. Nevertheless, the values of $q_{\rm EA}$ and $m$ can change indirectly through their dependence on the regular component $Q_{\rm reg}(\omega_k)$. In the following, we sketch the resulting saddle-point equations, and further technical details are provided in the Appendix \ref{app:derivation_saddle}.

The saddle point equation for $Q_{\rm reg}$ (see details in appendix \ref{app:derivation_saddle}) in the Matsubara frequency domain reads as follows:
\begin{align}\label{eq:Piregwn}
    \frac{\w^2_k}{\Gamma} + z &= \frac{1}{\qr(\w_k)} + \Pi_{\rm reg}(\w_k) 
\end{align}
where 
\begin{align}
    \Pi_{\rm reg}(\tau) &= \frac{3J^2}{2}\qd^2(\tau) - \frac{3J^2}{2}\qe^2 + V^2 \qd^3(\tau) G^4(\tau) 
\end{align}
The equation for decoupled bosons is obtained by setting $V=0$ in the above equation. Therefore, we see that the fermion-boson coupling gives a self-energy feedback to the bosons. In the later section, we will discuss the results for the effect of feedback in the characteristic feature of the regular part $Q_{\rm reg}(\tau)$. 

The saddle point equation for $q_{\rm EA}$ reads as below:
\beq 
	\frac{\qe}{\tilde{Q}_{aa}(0)^2-(m-1)\beta^2\qe^2+(m-2)\beta\tilde{Q}_{aa}(0)\qe} = \frac{ p}{2} \qe^{p-1} 
	\eeq 
This equation exactly reads same as the decoupled spherical boson model, but the $\tilde{Q}_{aa}(\w_k=0)$ gets affected by the fermion-boson feedback through Equation \ref{eq:Piregwn}.

The value of $m$ selects which metastable states we are considering:
{\it i)} when  $m$ is treated as a variational parameter that minimizes the free-energy functional $f(m)$, the target is the set of lowest metastable states dominating the Gibbs measure.

The boson–fermion coupling term 
\begin{align}
    \frac{NV^2}{4}\sum_{ab} \int d\tau d\tau' \, G^{ 4}_{ab}(\tau,\tau') \, Q^{ 4}_{ab}(\tau,\tau') \notag\\ 
\propto r \cdot \frac{NV^2}{4} \int d\tau d\tau' \, G^{ 4}(\tau,\tau') \, Q^{ 4}_{aa}(\tau,\tau')
\end{align}
reduces to a term proportional to the replica index $r$ because $G_{ab}(\tau) = G(\tau)\delta_{ab}$ is diagonal in replica space. As a result, this contribution does not depend  on $m$, and the variation of $f[m]$ reproduces the same set of equations as in the decoupled ($V=0$) $p$-spin-glass model. 

{\it ii)} if the breakpoint parameter $m$ is determined by imposing the marginality condition, requiring a second-order variation of the free-energy functional with respect to $Q_{ab}$ for $a \neq b$, then the highest `marginal' metastable states are selected. In this case again, the fermion-boson coupling term being diagonal in replica indices it does not change the marginality condition.

The set of equations which determine $m$ in thermodynamic spin-glass and marginal spin-glass phases are discussed in Appendix \ref{app:derivation_saddle}.

}

\subsubsection{Saddle point equation for pure SYK fermions vs fermions in coupled model}
Here, we briefly review the saddle-point equations for the pure SYK model, whose Hamiltonian is given by Eq.~\eqref{eq:SYKHam}. 
The coupling in the original SYK model has a random Gaussian-distribution with zero mean and finite standard deviation. In the large-$N$ limit, the Schwinger-Dyson (saddle-point) equations for SYK model take the following form:
\begin{align}
    G^{-1}(\w_n) &= -\ci\w_n - \Sigma(\w_n), \label{eq:Gwn} \\
    \Sigma(\tau) &= \mathcal{J}^2 G^3(\tau), \label{eq:sigmawn}
\end{align}
where $G(\tau)$ is the fermionic Green's function, and $\Sigma(\tau)$ is the self-energy. In the low-energy or long-time limit, the derivative term $(-\ci\w_n)$ can be neglected, leading to a conformal (power-law) solution. Moreover, the saddle-point equations exhibit an emergent time-reparameterization symmetry under arbitrary transformations $\tau \to f(\tau)$ in this limit.

In our model, the fermionic saddle-point equations, given by Eqs.~\eqref{eq:Gabwn} and~\eqref{eq:sigmaabwn}, can be comparable to the SYK saddle point equations~\ref{eq:Gwn} and ~\ref{eq:sigmawn} by interpreting an imaginary time dependent effective coupling as given by the form
\[
\mathcal{J}_{\rm eff}(\tau) = V \cdot Q^2_{aa}(\tau),
\]
where $Q_{aa}(\tau)$ is the diagonal part of bosonic Green's function. The effective $\mathcal{J}_{\rm eff}(\tau)$ encodes the fluctuating spin configuration of the quantum spin degrees of freedom. The explicit imaginary-time dependence of $\mathcal{J}_{\rm eff}(\tau)$ explicitly breaks the time-reparameterization symmetry and therefore does not support conformal solutions in general.  In the next section, we discuss the above time-dependent coupling in spin-glass phase in low temperatures can be effectively a static coupling, and we expect usual conformal invariant SYK Green's function in low temperatures.

\section{Coupled model: the spin-glass phase  }\label{sec:spinglass}

\begin{figure*}[htb!]
    \centering
\includegraphics[width=0.85\linewidth]{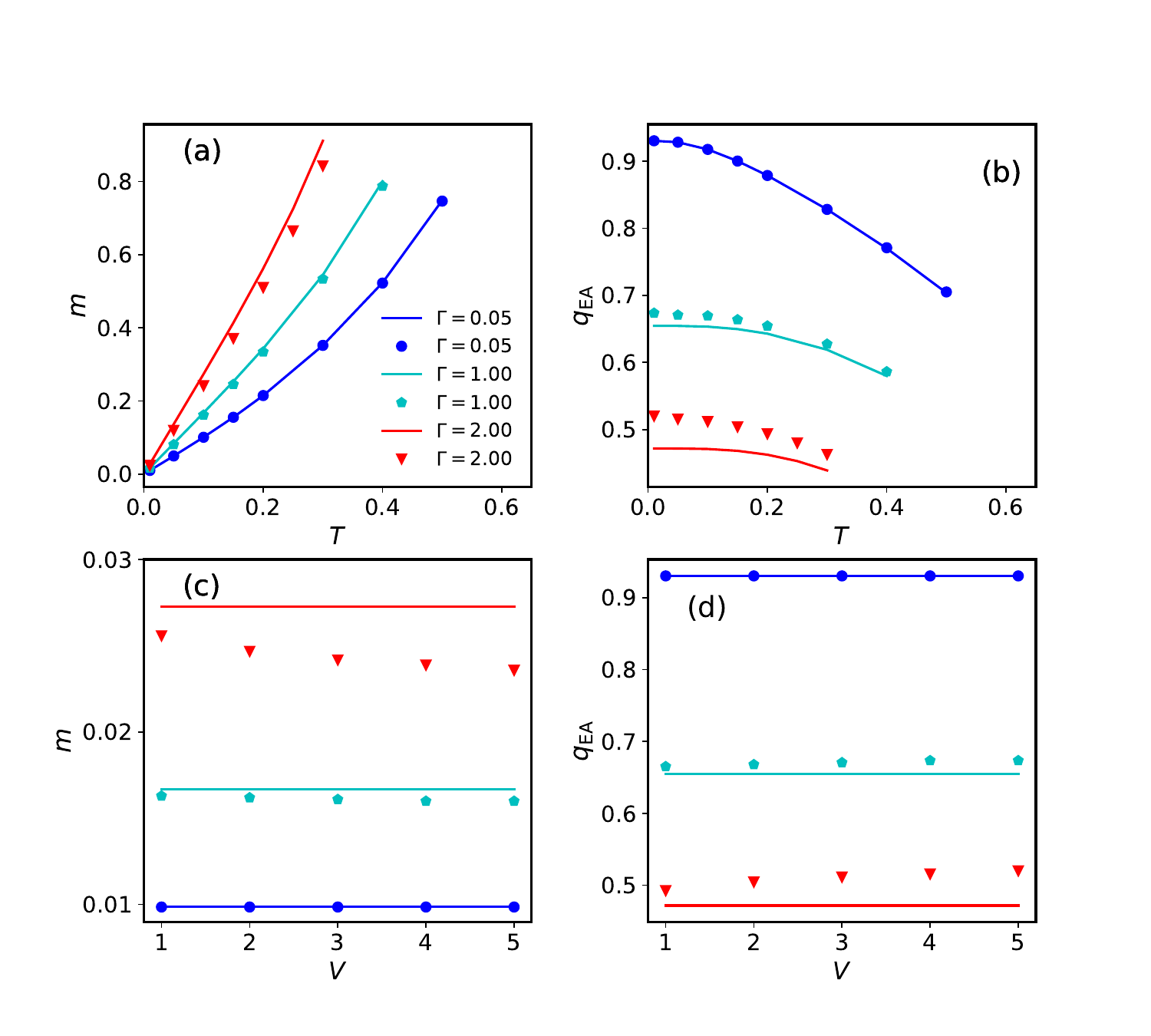}
 \caption{
\textbf{Effect of fermion-boson coupling on Edwards-Anderson and break-point parameter:} Panels (a) and (b) show the breakpoint parameter $m$ and the Edwards–Anderson parameter $q_{\rm EA}$, respectively, as functions of temperature $T$ for various values of the quantum parameter $\Gamma$, with the coupling strength fixed at $V/J = 5$. Solid lines correspond to the decoupled spin-glass system, while markers indicate results from the fully coupled boson–fermion system.
Panels (c) and (d) display $m$ and $q_{\rm EA}$, respectively, as functions of the boson–fermion coupling strength $V$ (with $J = 1$), for the same values of $\Gamma$ as in (a), at fixed temperature $T = 0.01$. As before, solid lines represent the decoupled case, and markers denote the fully coupled system.
}

    \label{fig:phasebnd_qe_m}
\end{figure*}

 In the glass phase we expect that the variables $\vec s$ are confined to pure states $\vec s^a$, where `$a$' labels the states. Each is present with
 a probability and an organization given by the Parisi ansatz. More generally, the system may find itself in metastable equilibrium in any metastable state, even those having relatively high free energies.
As we have mentioned above, within a state $a$, the fermions feel an effective interaction $\mathcal{J}^{(a)}_{mnpq} \equiv \sum_{ijkl} V^{mnpq} s_i^a s_j^a s_k^a s_l^a $. For low enough temperature and sufficiently low $\Gamma$ , one expects that each $\mathcal{J}^{(a)}_{mnpq}(\tau) \sim 
\sum_{ijkl} V^{mnpq} m_i^a m_j^a m_k^a m_l^a + 
\mathcal{J}^{fluctuating}_{mnpq}(\tau)$, where $m^a_i$ are the expectations of the $s_i$ within a state $a$ ($\sum (m_i^a)^2=q_{\rm EA}$). The fluctuating part has zero expectation and short time correlation, except for the marginal (threshold) states( See Figure \ref{fig:landscape}).

Consistently with the above picture, we find numerically that, in the classical limit $\Gamma \to 0$, the fermionic Green's function takes the form of that in the pure SYK model with a renormalized static coupling strength $\mathcal{J}^{(a)}_{\rm eff}$. Furthermore, the feedback effect of fermions on the bosonic self-energy is negligible; as a result, the bosonic Green's function remains unchanged upon coupling. 

In the following, we analyze the behavior of the bosonic and fermionic Green's functions in the glass phase for relatively large values of $\Gamma$. For strong quantum fluctuations ($\Gamma > 0$), tunneling between distinct pure states $a$ and $b$ can occur. At very low temperatures, the fermions predominantly experience the fluctuations of a single pure state $a$. At intermediate temperatures, however, where many (metastable) pure states become thermally accessible with appreciable probability within the Parisi hierarchical organization, the fermions effectively find a mixed environment arising from different pure states.

To this end, we present our results for the thermodynamic glass regime, followed by the results in the marginal glass phase. We begin by examining how the Edwards–Anderson parameter $q_{\rm EA}$ and the associated break-point parameter $m$ vary with temperature and coupling strength across different regions of the $\Gamma$–$T$ phase diagram. We then turn to the effect of boson–fermion coupling on both the bosonic Green's function $Q_{\rm reg}(\tau)$ (regular part)  and the fermionic Green's function $G(\tau)$.

\subsection{Effect of fermion-boson coupling on phase diagram}

First, we discuss the effect of fermion–boson coupling $(V)$ on the breakpoint parameter $m$ and the Edwards–Anderson parameter $q_{\rm EA}$. The breakpoint parameter $m$ takes values in the interval $0 \le m \le 1$. In the decoupled $p=3$ spin-glass model (i.e., for $V=0$), the phase boundary between the spin-glass and paramagnetic phases is determined by the condition $m=1$. This condition can be obtained from either the thermodynamic stability criterion or the marginal stability condition~\cite{Leticia2001}, as discussed in the previous section.
 In this phase transition, the Edwards–Anderson parameter $q_{\rm EA}$ shows a discontinuous jump from a finite value to zero. As shown in Refs.~\cite{Leticia2001, surajitPRL}, increasing quantum fluctuations (i.e., increasing $\Gamma$) tend to suppress $q_{\rm EA}$ and destabilize the spin glass phase at large $\Gamma$.

In Fig.~\ref{fig:phasebnd_qe_m}(a) and (b), we show the breakpoint parameter $m$ and the Edwards–Anderson parameter $q_{\rm EA}$, respectively, as functions of temperature $T$ for various values of the quantum parameter $\Gamma$, at fixed coupling strength $V/J = 5$. In Fig.~\ref{fig:phasebnd_qe_m}(c) and (d), we present the corresponding behavior as functions of the boson–fermion coupling strength $V$ (with $J = 1$), at fixed low temperature $T = 0.01$. In the Fig. \ref{fig:phasebnd_qe_m}, we determine $m$ and $q_{\rm EA}$ using the thermodynamic stability criteria. 

From these results, we observe that the coupling to fermions enhances $q_{\rm EA}$ and reduces the breakpoint parameter $m$. This effect becomes more pronounced at larger $\Gamma$, especially in the low-temperature regime. In the classical limit ($\Gamma \to 0$), we observe no change in either parameter. Since the phase boundary is defined by the condition $m = 1$, a reduction in $m$ implies that the transition line into the spin-glass phase shifts to higher temperatures. 
These findings indicate that coupling to fermions tends to stabilize and promote the spin-glass phase for large values of $\Gamma$. We find  similar qualitative results for marginal spin-glass phase.

\subsection{Green's function in thermodynamic spin-glass phase}

\begin{figure*}[htb]
    \centering
\includegraphics[width=0.85\linewidth]{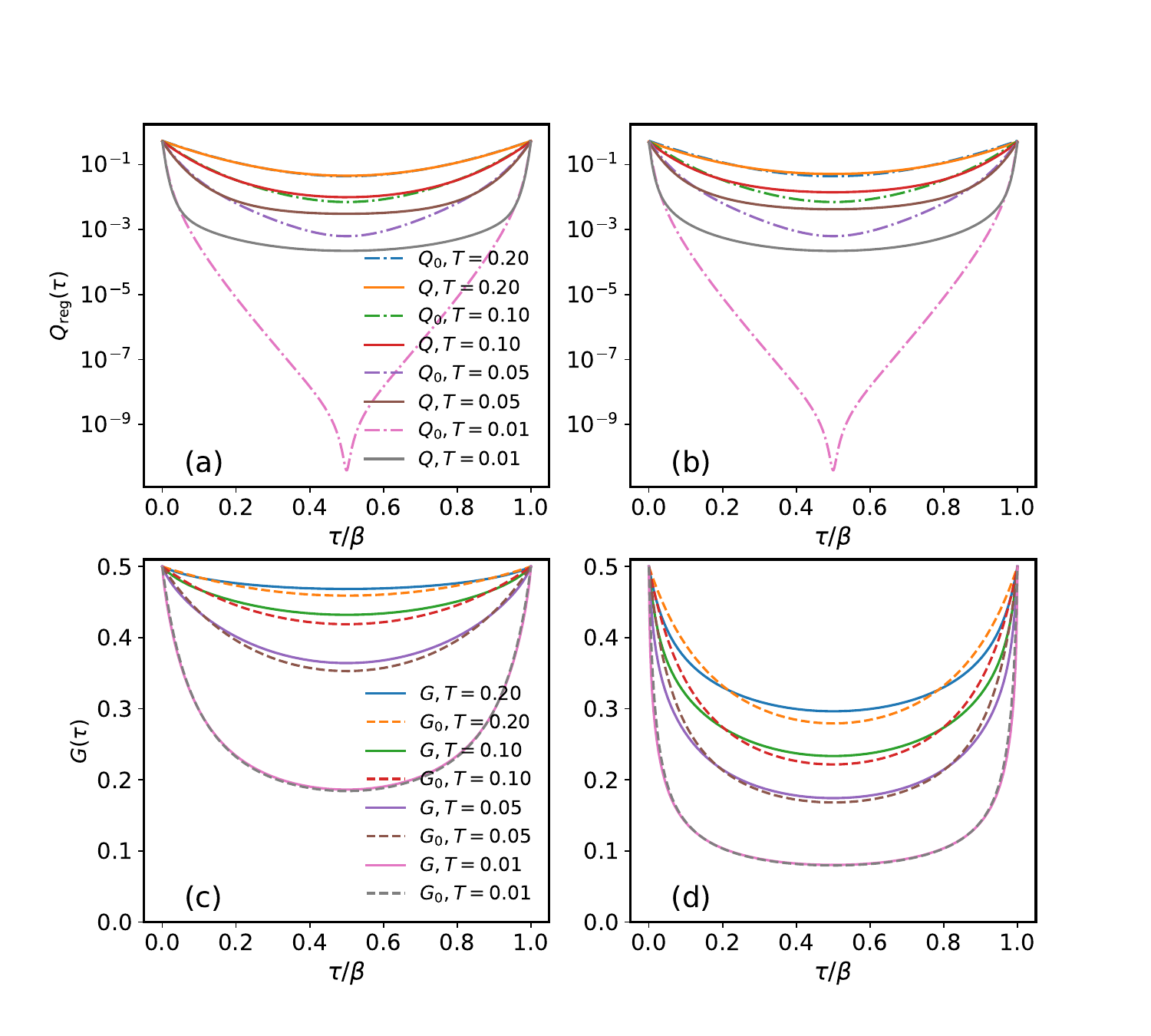}
\caption{
\textbf{Upper panel (a, b) -- Bosonic Green's function in thermodynamic spin-glass phase:} The regular part of the bosonic Green's function, defined as $Q_{\rm reg}(\tau) = Q_{aa}(\tau) - q_{\rm EA}$, is plotted as a function of $\tau/\beta$ for various temperatures $T$. Panel (a) corresponds to quantum fluctuation parameter $\Gamma = 2.0$ and coupling strength $V/J = 1.0$, while panel (b) shows results for $V/J = 5.0$. In the legend, $Q_0$ denotes the bosonic Green's function in the absence of fermion–boson coupling ($V = 0$), and $Q$ corresponds to the fully coupled case ($V \ne 0$). 
\textbf{Lower panel (c, d) -- Fermionic Green's function:} The corresponding fermionic Green's functions are shown in panels (c) and (d), with parameters matching those in (a) and (b), respectively. In the legend, $G$ represents the fermionic Green's function of the fully coupled boson–fermion system, and $G_0$ refers to the SYK Green's function computed with a renormalized effective coupling $\mathcal{J}_{\rm eff} = V \cdot \langle Q^2(\tau) \rangle$.
}

    \label{fig:Q-tSG-J2p0}
\end{figure*}

\begin{figure}[htb]
    \centering
    \includegraphics[width=0.99\linewidth]{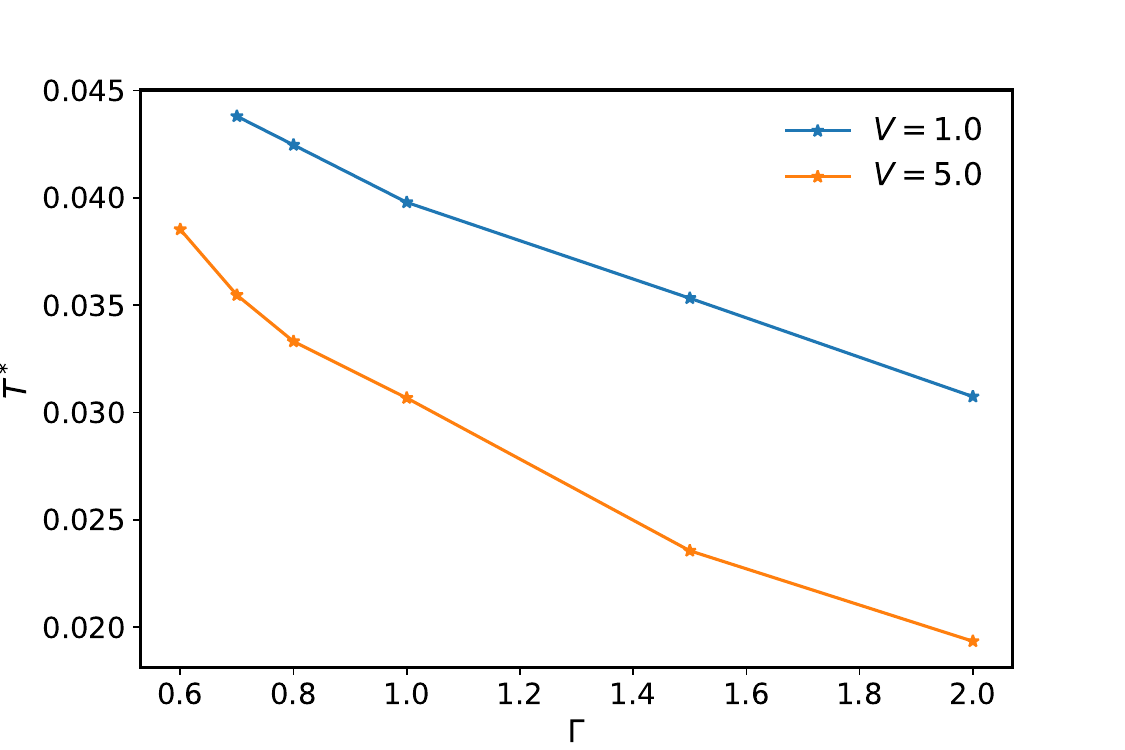}
\caption{Crossover temperature $T^*(\Gamma,V)$ as a function of $\Gamma$ for several coupling strengths $V$, characterizing the departure from SYK behavior in the thermodynamic spin-glass phase.}

    \label{fig:crossoverT}
\end{figure}

\begin{figure*}[htb]
    \centering
\includegraphics[width=0.8\linewidth]{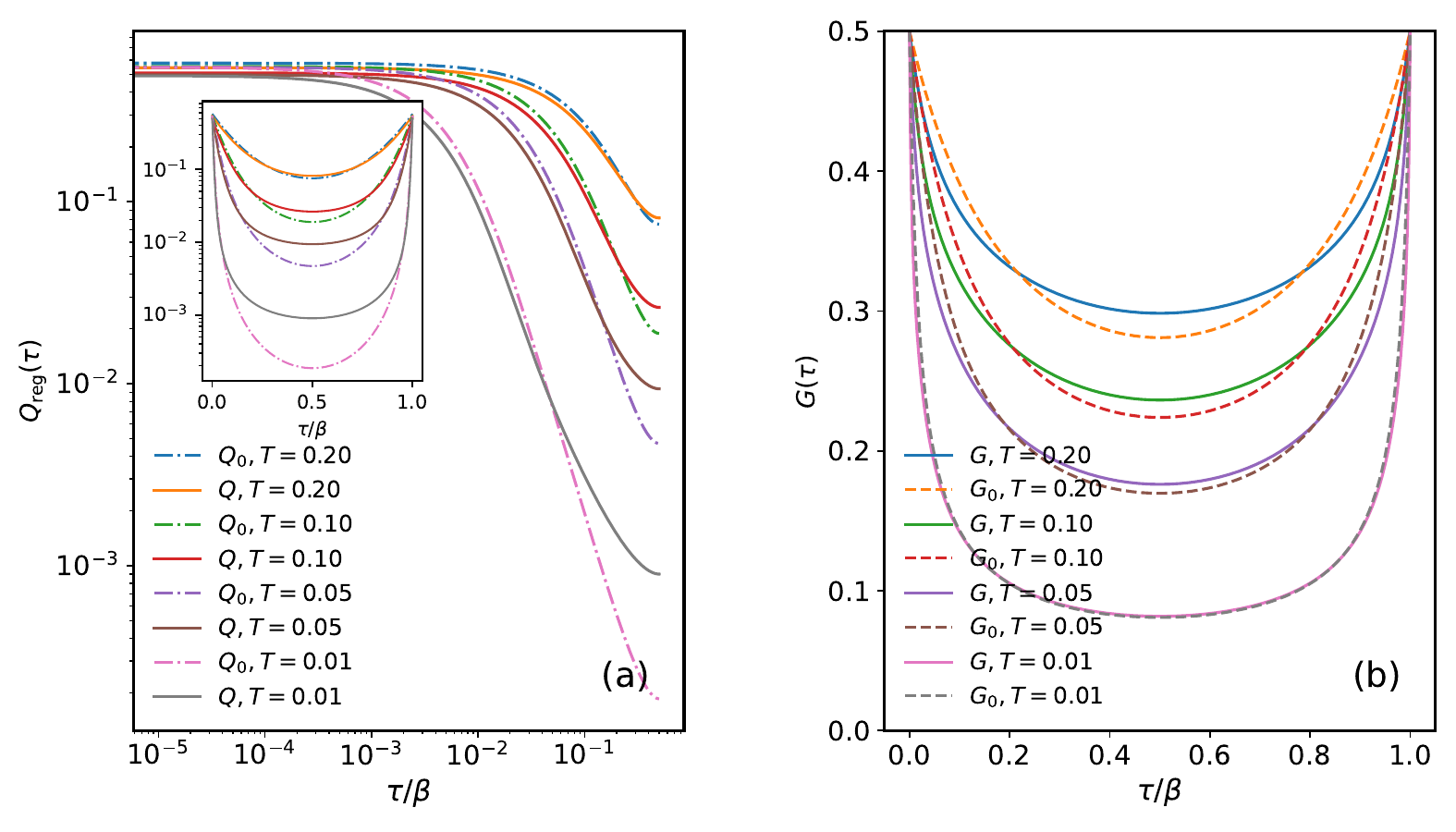}
\caption{
\textbf{Left panel (a) -- Bosonic Green's function in marginal spin-glass phase:} The regular part of the bosonic Green's function is plotted as a function of $\tau/\beta$ (upto $\tau=\beta/2$) for different $T$ in log-log scale in panel (a). The result corresponds to $\Gamma = 2.0$ and coupling strength $V/J = 5.0$. In the legend, $Q_0$ denotes the bosonic Green's function in the absence of fermion–boson coupling ($V = 0$), and $Q$ corresponds to the fully coupled case ($V \ne 0$).  In the inset, the results are shown in semi-logarithmic scale (y-axis) to show the full $Q(\tau)$ vs $\tau/\beta$ profile.
\textbf{Right panel (b) -- Fermionic Green's function :} The corresponding fermionic Green's functions are shown in panel (b) with parameters matching those in (a). In the legend, $G$ represents the fermionic Green's function of the fully coupled boson–fermion system, and $G_0$ refers to the pure SYK Green's function computed with a static renormalized effective coupling $\mathcal{J}_{\rm eff} = V \cdot \langle Q^2(\tau) \rangle$.
}

    \label{fig:Q-mSG-J2p0}
\end{figure*}

In Figure~\ref{fig:Q-tSG-J2p0}(a), (b), we plot the regular part of the bosonic Green's function, $Q_{\rm reg}(\tau) = Q_{aa}(\tau) - q_{\rm EA}$ for various values of temperature $T$. Panels (a) and (b) correspond to boson–fermion coupling strengths $V/J = 1.0$ and $V/J = 5.0$, respectively for quantum parameter $\Gamma=2.0$. We obtain the result here using the thermodynamic spin-glass criterion. Each plots include both the Green's function for the decoupled case ($V = 0$), labeled as $Q_0$, and the fully coupled case, labeled as $Q$. The results are shown on a normal-log scale, with a linear $x$-axis and logarithmic $y$-axis, which allows us to examine the exponential decay behavior of the Green's function.

We observe that $Q_{0, \rm reg}(\tau)$ exhibits a linear profile on the semi-logarithmic scale over the interval $\tau \in [0, \beta/2)$, indicating an exponential decay of the form $Q_{0,\rm reg}(\tau) \sim e^{-E_g \tau}$ in this regime. This behavior show the presence of an energy gap $E_g$ in the equilibrium spin-glass excitation spectrum, and consequently the low-temperature specific heat scaling $C_v \sim e^{-E_g/T}$\cite{Leticia2001}.

The effect of coupling on the bosonic Green's function can be summarized as follows. First, the value of $Q_{\rm reg}(\tau)$ near $\tau \sim \beta/2$ is significantly altered upon coupling to fermions, with the modification becoming more pronounced at lower temperatures. Second, the exponential decay of $Q_{\rm reg}(\tau)$ observed in the decoupled case over the interval $\tau \in [0, \beta/2)$ is replaced in the coupled system by a plateau--like behavior at intermediate times, characterized by much slower temporal variation. This qualitative change -- from exponential decay to an approximate plateau -- is induced by the fermionic self-energy feedback, which modifies the long-time structure of the bosonic Green's function and consequently changes the gap excitation spectrum of the equilibrium spin-glass phase. We note that this effect is already present for $V/J=1$ (see panel (a)), increasing the coupling $V/J$ mainly affects the range of temperatures where deviations become important.

In Figure~\ref{fig:Q-tSG-J2p0}(c) and (d), we show the fermionic Green's function $G(\tau)$ for various temperatures with the parameters same as those used in Figure~\ref{fig:Q-tSG-J2p0}(a) and (b) respectively. All plots are shown on a linear scale. The solid lines represent results from the fully coupled fermion–boson system, while the dotted lines correspond to the pure SYK Green’s function computed with an effective renormalized static coupling $\mathcal{J}_{\rm eff} = V \cdot \langle Q^2(\tau) \rangle$, where the bosonic Green’s function $Q(\tau)$ is replaced by its average value in the fermionic self-energy.

At very low temperatures, such as $T = 0.01$, the fermionic Green’s function matches precisely with the SYK Green’s function using the static renormalized coupling $\mathcal{J}_{\rm eff}= V \cdot \langle Q^2(\tau) \rangle$. This agreement can be attributed to the bosonic Green’s function becoming nearly time-independent at low temperatures—i.e., $Q(\tau) \approx q_{\rm EA} + Q_{\rm reg}(\tau)$ with $Q_{\rm reg}(\tau)$ nearly constant—making the replacement by a constant average value a good approximation.

However, as the temperature increases, deviations between the full fermionic Green’s function and the SYK Green's function with a renormalized static coupling become noticeable. This discrepancy is more pronounced at higher coupling strengths (e.g., $V/J = 5.0$). This behavior can be understood within the picture of pure-state organization in the Parisi scheme described earlier. At very low temperatures, the bosonic spins are confined to a specific pure state $a$, so the fermions effectively experience a static background interaction characterized by $\mathcal{J}^{(a)}_{\rm eff}$. In contrast, at intermediate temperatures, other metastable states acquire non-negligible statistical weight, and tunneling between different pure states in the spin-glass landscape becomes likely. Consequently, the fermions experience a state-mixing dynamic $\mathcal{J}_{\rm eff}(\tau)$ coupling  environment, which leads to deviations from the pure SYK behavior with a static $\mathcal{J}_{\rm eff}= V \cdot \langle Q^2(\tau) \rangle$.
This suggests the existence of a crossover temperature $T^*(\Gamma, V)$ separating a low temperature regime where the SYK fixed point is recovered from an intermediate one where significant deviations appear. 

{We present the crossover temperature $T^*(\Gamma,V)$ as a function of $\Gamma$ for several values of the coupling strength $V$ in Fig.\ref{fig:crossoverT}. To determine $T^*(\Gamma,V)$, we compute the absolute difference between the fully interacting numerical fermionic Green’s function $G(\tau)$ and the SYK Green’s function $G_0(\tau)$ evaluated with a static effective coupling $\mathcal{J}_{\rm eff}= V\,\langle Q^2(\tau)\rangle$. This difference is then compared against an empirically chosen threshold, as discussed in detail in the Appendix \ref{Appendix:CrossoverT}. We find that $T^*(\Gamma,V)$ shifts towards lower values as the interaction strength $V$ is increased. This effect becomes more pronounced for large values of the quantum  parameter $\Gamma\gg 1$, corresponding to the strongly quantum regime. In contrast, for small $\Gamma\lesssim 0.5$, i.e., in the classical limit, we find that the solution remains SYK-like over the entire accessible temperature range when described by a static effective coupling. This behavior is consistent with the physical picture discussed in earlier sections.
}

{
In the intermediate regime, for example at $\Gamma=0.6$ and $0.7$, we observe a nontrivial temperature dependence. At very low temperatures, the fermionic Green’s function closely follows the SYK solution. Upon increasing temperature, the system undergoes a crossover to a non-SYK regime, while at sufficiently high temperatures the Green’s function again agrees with SYK behavior, as shown in Fig.~\ref{fig:TVD_vs_T} in the Appendix \ref{Appendix:CrossoverT}. This re-entrant SYK behavior at high temperatures can be understood from the fact that, in this regime, the bosonic correlator $Q(\tau)$ becomes only weakly time-dependent. Consequently, the effective coupling $\mathcal{J}_{\rm eff}(\tau)=V\cdot\,Q_{aa}^2(\tau)$ can once again be approximated by a static interaction for the fermions.
}

Moreover, in Appendix~\ref{sec:scaling-SYK-tSG}, we also present the power-law scaling of the fermionic Green’s function and demonstrate how it is gradually modified as the temperature is increased to intermediate values.

\subsection{Green's function in marginal spin-glass phase}

Here, we present the results for the bosonic Green's function in the coupled problem, obtained under the marginal stability criterion. Marginal stability requires the replicon eigenvalue to vanish, selecting a flat direction in the free-energy functional, as illustrated in Fig.~\ref{fig:landscape}. In the decoupled $p$-spin glass model under this condition, the excitation spectrum is gapless critical behaviour and the regular part of the correlator exhibits a power-law decay, $Q_{0,\mathrm{reg}}(\tau) \sim 1/\tau^{\alpha}$~\cite{Leticia2001}. Consequently, the specific heat displays linear temperature dependence, $C_V \propto T$~\cite{Leticia2001}. 

Figure~\ref{fig:Q-mSG-J2p0}(a) shows the bosonic Green's function $Q_{\mathrm{reg}}(\tau)$ for the coupled system and $Q_{0,\mathrm{reg}}(\tau)$ for the decoupled case as functions of $\tau/\beta$ up to $\tau = \beta/2$, plotted on a log–log scale. The inset presents the full profiles of both correlators on a semi-logarithmic scale (logarithmic $y$-axis). These results are shown for $V/J = 5.0$ and $\Gamma = 2.0$. As is evident from the figure, coupling to fermions shortens the power-law regime (linear segment in the log–log plot) at $\tau < \beta/2$ and leads to the development of a slowly varying plateau near $\tau \sim \beta/2$. 

The corresponding fermionic Green's function for the coupled system, along with its comparison to the pure SYK Green's function with a static renormalized coupling, is shown in Fig.~\ref{fig:Q-mSG-J2p0}(b). Its behavior closely parallels that observed in Fig.~\ref{fig:Q-tSG-J2p0}(d) for the thermodynamic glass phase.

\section{ Coupled model: paramagnetic phase }\label{sec:PMphase}
In this section, we discuss how the fermion-boson coupling in the paramagnetic phase affects the bosonic and fermion Green's function.

\begin{figure*}[htb]
    \centering
\includegraphics[width=0.85\linewidth]{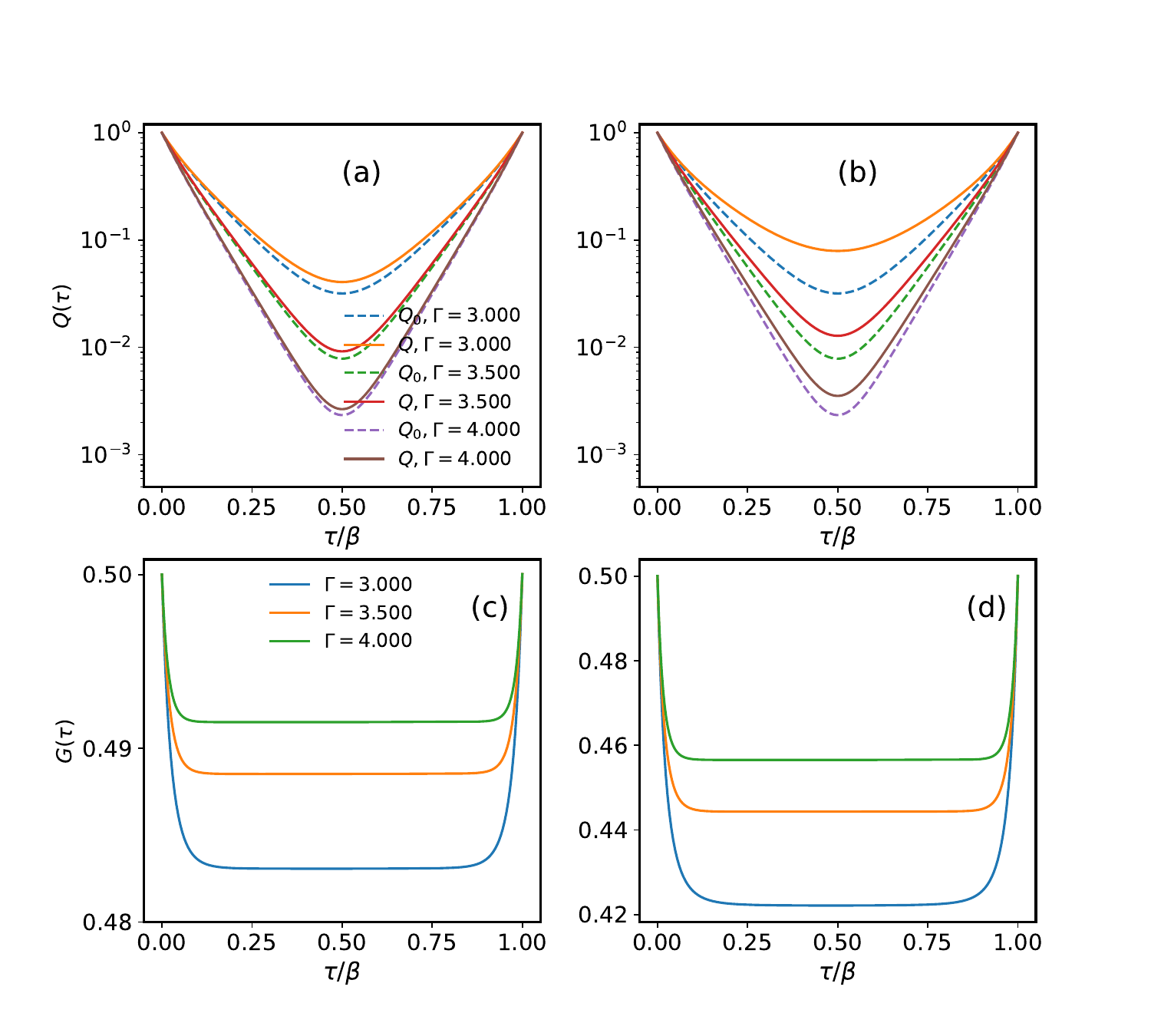}
\caption{
\textbf{Upper panel (a, b) -- Bosonic Green's function for quantum paramagnet:} The bosonic Green's function $Q(\tau)$ is shown for several values of the quantum fluctuation parameter $\Gamma$ in the quantum paramagnetic phase at $T=0.125$. Panel (a) corresponds to boson–fermion coupling strength $V/J = 2.0$, while panel (b) shows results for $V/J = 5.0$. The curve labeled $Q_0$ represents the bosonic Green's function in the decoupled case ($V = 0$). 
\textbf{Lower panel (c, d) -- Fermionic Green's function:} The corresponding fermionic Green's functions for the coupled boson–fermion system are displayed in panels (c) and (d), using the same parameters as in panels (a) and (b), respectively. 
}

    \label{fig:Q-qPM}
\end{figure*}

\subsection{Quantum paramagnet}

In Figure~\ref{fig:Q-qPM} (a) and (b), we present the bosonic Green's function $Q(\tau)$ for the coupled system ($V \neq 0$) at several values of the quantum parameter $\Gamma = 3.0,\ 3.5,\ 4.0$ corresponding to the quantum paramagnetic solutions. Panels (a) and (b) correspond to boson–fermion coupling strengths $V/J = 2.0$ and $V/J = 5.0$, respectively, computed at temperature $T=0.125$. These plots are presented on a semi-logarithmic scale with logarithmic y-axis. In each case, we compare the results with those from the decoupled system ($V = 0$), denoted as $Q_0(\tau)$. 

In the quantum paramagnetic phase, the bosonic Green's function exhibits exponential decay over the interval $\tau \in [0, \beta/2]$ ($Q(\tau<\beta/2)\propto e^{-\Delta\tau}$, $\Delta$ is the decay rate)), as evidenced by the clear linear behavior in the semi-logarithmic plots. This exponential decay corresponds to a gapped spectral function in the real-frequency domain. Upon coupling to fermions, we observe that the value of $Q(\tau)$ near $\tau \sim \beta/2$ is only mildly affected, and the exponential decay characteristic of the quantum paramagnetic phase is largely preserved. 

The corresponding fermionic Green's functions for the same parameters as in Figure~\ref{fig:Q-qPM} (a) and (b) are shown in panels (c) and (d), respectively, plotted on a linear scale. We find that the fermionic Green's function exhibits a rapid initial decay followed by a plateau at intermediate times, with $G(\tau \sim \beta/2) < 0.5$. The plateau value decreases with increasing coupling strength $V/J$, and for a fixed $V/J$, it further decreases with decreasing $\Gamma$.

{We emphasize that this result is rather surprising, since one might have naively expected that coupling the SYK fermions to a gapped phase, such as the quantum paramagnet, would not significantly modify the fermionic dynamics. Instead, our results demonstrate that the parametric coupling between fermions and gapped bosons can induce a qualitative restructuring of the fermionic Green’s function.}

{The effective time-dependent coupling experienced by the fermions,
\[
\mathcal{J}_{\rm eff}(\tau)=V\,Q^2(\tau),
\]
 becomes exponentially suppressed at long imaginary times as $Q(\tau)$ being exponential as shown in Fig. \ref{fig:Q-qPM}. One might therefore expect that the resulting fermionic dynamics would be similar to that of the SYK model with vanishing interaction strength.
}

{
However, the nature of the emergent slow fermionic dynamics differs markedly from the $\mathcal{J}\to 0$ limit of the pure SYK model. For completeness, the imaginary-time Green’s function of the SYK model in the $\mathcal{J}\to 0$ limit is presented in Appendix~\ref{sec:G0-tau}. We also observe that the interacting Green’s function obtained here does not reduce to that of free Majorana fermions, for which
\begin{equation}\label{eq:freeMajGreen}
G(\tau)=\frac{1}{2}\,\mathrm{sgn}(\tau).
\end{equation}
}
{
These observations suggest that the effect of the interaction can rather be a perturbative correction to free fermion propagator at long imginary time (low-energy limit). Although the bare fermion--boson coupling $V$ may be large, the effective interaction experienced by the fermions becomes parametrically small at long imaginary times due to the exponential decay of the bosonic propagator. To gain analytical insight into the low-energy behavior, we therefore consider the zero-temperature limit of the Schwinger--Dyson equations and perform a first-order perturbative expansion of the fermionic self-energy.
}

{
As in this quantum paramagnetic regime, the bosonic Green’s function retains its qualitative exponential decay, we approximate the $T=0$ bosonic correlator as
\begin{equation}
Q_0(\tau)=e^{-\Delta|\tau|},
\end{equation}
where $\Delta$ denotes the bosonic gap controlling the decay rate. This form satisfies the spherical constraint $Q_0(\tau=0)=1$ and is symmetric under $\tau\to -\tau$. We then evaluate the fermionic self-energy perturbatively using the free Majorana propagator Eqn. \ref{eq:freeMajGreen},
}

{
At leading order, the fermionic self-energy takes the form (see Appendix \ref{app:pert_self_energy_qpm} for details)
\begin{equation}
\Sigma^{(1)}(\tau)
=
\frac{V^2}{8}\,e^{-4\Delta|\tau|}\,\mathrm{sgn}(\tau).
\end{equation}
The corresponding first-order corrected fermionic Green’s function in frequency space is given by
\begin{align}
G^{(1)}(i\omega)
&=
\frac{1}{-i\omega}\,
\frac{1}{1+\dfrac{V^2}{4(\omega^2+16\Delta^2)}} \nonumber\\
&\simeq
\frac{1}{-i\omega}
\left[
1-\frac{V^2}{4(\omega^2+16\Delta^2)}+O(V^4)
\right].
\end{align}
}

{
In the low-energy limit $|\omega|\ll\Delta$, this expression simplifies, and transforming back to imaginary time yields
\begin{equation}
G^{(1)}(\tau)
\simeq
\frac{1}{2}
\left(
1-\frac{V^2}{64\,\Delta^2(\Gamma)}
\right)
\mathrm{sgn}(\tau).
\end{equation}
\\
Although this perturbative result is derived strictly at $T=0$ and is not expected to hold quantitatively at finite temperature, it nevertheless captures the essential low-energy physics observed in our numerical solutions. In particular, it explains the reduction of the plateau value of the fermionic Green’s function below $1/2$ and its systematic decrease with increasing coupling strength $V$. Moreover, this effect is enhanced at smaller $\Gamma$, for which the bosonic gap satisfies $\Delta(\Gamma_1)<\Delta(\Gamma_2)$ when $\Gamma_1<\Gamma_2$.
\\
\\
We therefore conclude that the plateau observed in the fermionic Green’s function arises as a low-energy perturbative correction to the free-fermion propagator induced by coupling to an exponentially decaying, gapped bosonic environment. Within the above perturbative treatment, we are unable to capture the initial fast decay of the fermionic Green’s function, as this regime is controlled by high-frequency contributions that lie beyond the low-energy approximation employed here.}

 \begin{figure*}[htb!]
    \centering
\includegraphics[width=0.85\linewidth]{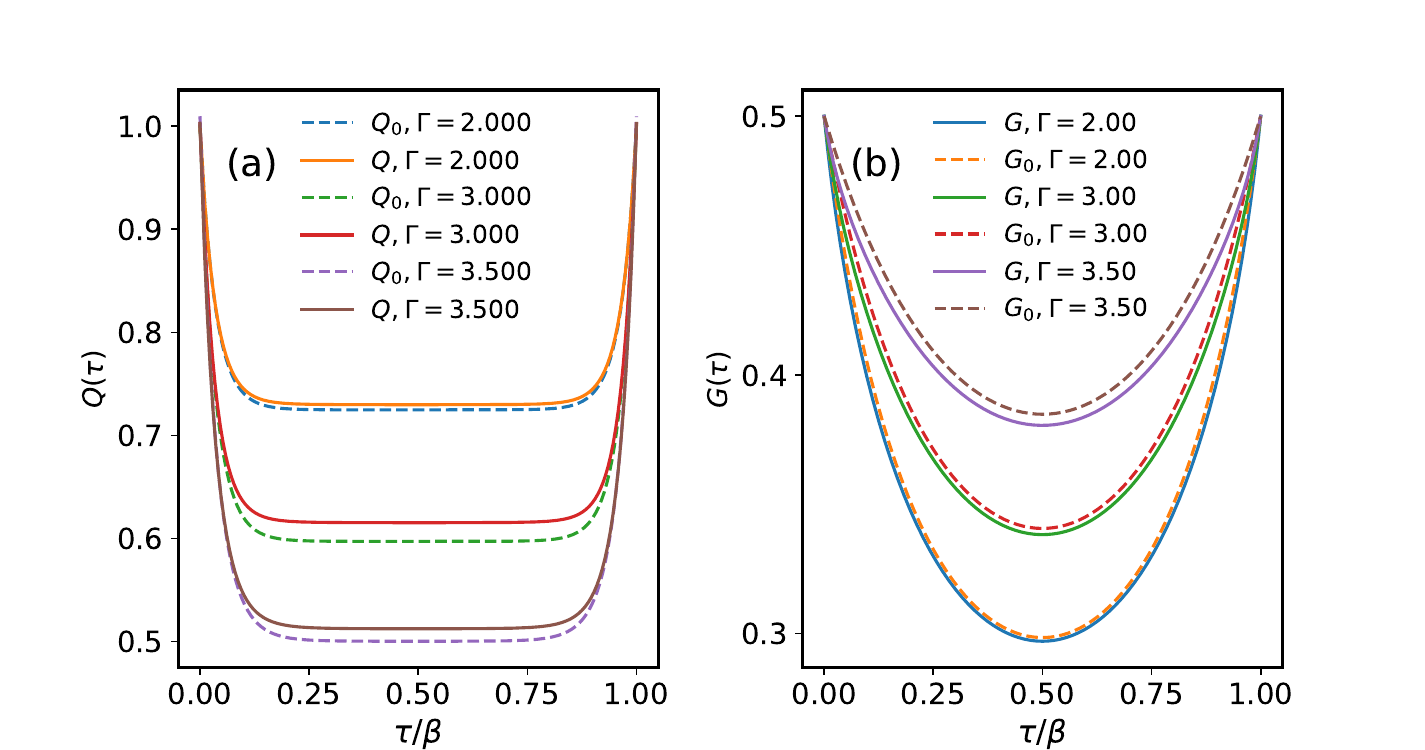}
 \caption{ (a) \textbf{Bosonic Green's function for classical paramagnet: }
 The classical paramagnetic solutions are shown for different values of the quantum parameter $\Gamma$ at fixed temperature $T = 0.125$. Both the decoupled ($Q_0$) and coupled ($Q$) bosonic Green's functions are plotted for comparison. The boson–fermion coupling strength is set to $V/J = 2.0$. 
(b) \textbf{Fermionic Green's function: }  The corresponding fermionic Green's functions are displayed. Here, $G$ represents the full numerical solution for the coupled system, while $G_0$ denotes the pure SYK Green's function with an effective coupling $J_{\rm eff} = V \cdot q_P^2$, where $q_P$ is the plateau value of $Q(\tau)$ from panel (a).
}

    \label{fig:c-PM}
\end{figure*}

\subsection{Classical paramagnet}

In Figure~\ref{fig:c-PM} (a), we present the bosonic Green's function $Q(\tau)$ for several values of the quantum parameter $\Gamma = 2.0,\ 3.0,\ 3.5$ corresponding to classical paramagnetic solutions at temperature $T=0.125$. The plot is shown on a linear scale, and the boson–fermion coupling strength is fixed at $V/J = 2.0$. For comparison, we also show the decoupled case ($V = 0$), denoted as $Q_0(\tau)$. These solutions are termed \textit{classical paramagnetic} since the mid-point value $Q(\tau = \beta/2)$ is much larger than in the quantum paramagnetic solutions presented in Figure~\ref{fig:Q-qPM}. The classical paramagnet is characterized by an early decay in $\tau$, followed by a plateau at value $q_P$. The coupling to fermions has only a minimal effect on the plateau value $q_P$ for $V/J = 2.0$, and the deviation becomes more noticeable at stronger coupling (result not presented here). 

The corresponding fermionic Green's functions $G(\tau)$ are displayed in Figure~\ref{fig:c-PM} (b) for the same parameters as in panel (a). The fermionic Green's function closely matches the pure SYK Green's function computed using an effective coupling $\mathcal{J}_{\rm eff} = V \cdot q_P^2$, where $q_P$ is the plateau value of $Q(\tau)$ from panel (a). This indicates that, in the classical paramagnetic regime, coupling to fermions does not qualitatively alter the behavior of either the bosonic or fermionic Green's functions.
 
\section{Conclusion and Discussion}
\label{sec:discussion}

In this work, we studied a model of disordered fermion–boson interactions, where Majorana fermions interact via all-to-all random four-fermion couplings as in the SYK model, and are simultaneously coupled to bosonic degrees of freedom governed by a quantum spherical $p=3$ spin-glass model. The bosonic sector features a rugged free energy landscape and exhibits a spin-glass phase at low temperatures characterized by a one-step replica symmetry breaking (1-RSB) structure, and a paramagnetic phase at large quantum fluctuations ($\Gamma \gg 1$) or high temperature.

The fermions, evolving within the valleys of this complex landscape, experience effective couplings that depend sensitively on the organization of pure states in the Parisi scheme. When the bosonic density matrix is dominated by a single pure state, the fermions effectively experience a static, renormalized coupling of the form $\mathcal{J}^{(a)}_{mnpq} = \sum_{ijkl} V^{mnpq}_{ijkl} s^{(a)}_i s^{(a)}_j s^{(a)}_k s^{(a)}_l$. In contrast, when multiple pure states contribute significantly, the fermions encounter a dynamic, state-mixing coupling in imaginary time. Our large-$N$ saddle-point analysis supports this intuitive picture: in the spin-glass phase, the fermionic Green’s function follows the SYK form at low temperatures, while deviations from this behavior appear at higher temperatures due to dynamical mixing.

The feedback of fermions on the bosonic sector is equally striking. In the decoupled limit, the bosonic Green’s function decays exponentially in the equilibrium spin-glass phase and as a power law in the marginal spin-glass regime. Upon coupling to fermions, both behaviors are replaced by a slowly varying, plateau-like profile in imaginary time, signaling a qualitative shift in bosonic dynamics. In the paramagnetic phase, however, the bosonic Green’s function remains largely unaffected—retaining its plateau-like structure in the classical regime ($\Gamma \ll 1$) and exponential decay in the quantum regime ($\Gamma \gg 1$). The fermionic sector, by contrast, displays rich behavior: in the classical paramagnetic regime, it remains well described by the SYK form upon coupling, whereas in the quantum paramagnetic regime, it departs significantly from the SYK behavior and instead develops a plateau-like structure.

Overall, our results demonstrate how disordered fermion–boson coupling qualitatively modifies both bosonic and fermionic correlators, revealing a crossover from SYK-like behavior at low temperatures to distinctly non-SYK behavior in the paramagnetic regime dominated by strong quantum fluctuations.

An interesting direction for future exploration concerns the out-of-equilibrium dynamics of this coupled system. The relaxation dynamics of quantum glasses has been extensively studied both in presence of dissipation and under unitary dynamics~\cite{cugliandolo1999realtime,cugliandolo2002dissipative,Cugliandolo_2019,thomson2020quantum,lang2024replica,lang2025numericalrenormalizationglassydynamics}. When the bosonic sector is connected to a zero-temperature bath, it is known to exhibit aging: the system evolves increasingly slowly without ever equilibrating. In this setting, the effective fermionic couplings $\mathcal{J}_{mnpq}$ become slowly time-dependent, mirroring the aging of the bosons. Since the bosonic fluctuations are governed by an effective temperature $T_{\mathrm{eff}} > 0$ \cite{LeticiaGustavo}, the fermions inherit this effective temperature and never fully reach a zero-temperature state as long as the system remains out of equilibrium. This behavior would be therefore in stark contrast with the dynamics of a pure SYK model which instead displays fast thermalization~\cite{eberlein2017quantum,bhattacharya2019quantum,haldar2020quench,quantum2021cheipesh,larzul2022quenches,jaramillo2025thermalization}.

Understanding the consequences of such aging dynamics on the fermionic sector and more broadly, on the interplay between quantum chaos and glassiness remains a compelling question for future study.

\section{Acknowledgements}
S.B. and M.S. acknowledge funding from the European Research Council (ERC) under the European Union's Horizon 2020 research and innovation program (Grant agreement No. 101002955 -- CONQUER). We acknowledge Coll\`{e}ge de France IPH cluster where the numerical calculations were carried out.

\appendix
\onecolumngrid

\section{Derivation of saddle point equations}\label{app:derivation_saddle}
To compute the disorder average free energy  $F=-k_B T \overline{\log Z}$, we use the replica trick as follows:
\begin{align}
    \log Z &= \lim_{r\to 0} \frac{Z^r-1}{r}
\end{align}
The replicated partition function of the Hamiltonian in eqn. \ref{eqn:full-hamiltonian} is given as follow:
\begin{align}
    Z^r &= \int \mathcal{D}[s_a,\chi_a] e^{-S}
\end{align}
where the imaginary time action $S$ is given by:
\begin{align}
    S &= \frac{1}{2}\int d\tau \sum_{m,a} 
 \chi_{ma}(\tau)\partial_{\tau}\chi_{ma}(\tau) +  \int d\tau \bigg(\sum_{i,a}\frac{M}{2}\left(\frac{\partial s_{ia}}{\partial \tau }\right)^2 + \sum_{ijk, a}J_{ijk}s_{ia}(\tau)s_{ja}(\tau)s_{ka}(\tau)\bigg) \notag\\
&+ \int d\tau \sum_{ijkl;mnpq,a} V_{ijkl;mnpq} s_{ia}(\tau)s_{ja}(\tau)s_{ka}(\tau)s_{la}(\tau) \chi_{ma}(\tau)\chi_{na}(\tau)\chi_{pa}(\tau)\chi_{qa}(\tau)
\end{align}
In the above $a\in [1, r]$ is replica indices. In the end we take analytical continuation of $r\to 0$. Performing disorder average on $V_{ijkl;mnpq}$ yields:
\begin{align}
\prod^{m<n<p<q}_{i<j<k<l} \int dV_{ijkl;mnpq} \exp\bigg[-\frac{V^2_{ijkl;mnpq}}{2\sigma^2} - V_{ijkl;mnpq}\int d\tau\sum_a s_{ia}...s_{la}\chi_{ma}...\chi_{qa}  \bigg] \notag\\
= \exp\bigg[ \frac{NV^2}{4}\int d\tau d\tau' \sum_{ab} \bigg(\frac{1}{N}\sum_{i}s_{ia}(\tau)s_{ib}(\tau')\bigg)^4\bigg(\frac{1}{N}\sum_{m}\chi_{ma}(\tau)\chi_{mb}(\tau') \bigg)^4\bigg]
\end{align}
Similarly, we can perform the disorder average for $J_{ijk}$ for $p=3$ spin-glass part and we obtain:
\begin{align}
 \exp\bigg[-\int d\tau\sum_{ia} s_{ia}\big(-\frac{M}{2}\partial^2_{\tau} + \frac{z}{2}\big)s_{ia}(\tau)+ \frac{NJ^2}{4}\int d\tau d\tau' \sum_{ab}\bigg(\frac{1}{N}\sum_{i}s_{ia}(\tau)s_{ib}(\tau')\bigg)^3 \bigg]
\end{align}
where $z$ is the Lagrange multiplier to enforce the spherical constraint. 
We introduce following large-$N$ dynamical fields
\begin{align}
    G_{ab}(\tau,\tau') &= \frac{1}{N} \sum_{i}\chi_{ia}(\tau)\chi_{ib}(\tau')\\
 Q_{ab}(\tau,\tau') &= \frac{1}{N}\sum_{i}s_{ia}(\tau)s_{ib}(\tau') 
\end{align}
and we obtain disorder-average replicated partition function in terms of the above dynamical fields as follows:
\begin{align}
\overline{Z^r} &= \int \mathcal{D}[s,\chi] \exp\bigg[-\int d\tau \frac{1}{2}\sum_{m, a} 
 \chi_{ma}(\tau)\partial_{\tau}\chi_{ma}(\tau)+\frac{NV^2}{4}\int d\tau d\tau' \sum_{ab} Q^{\circ 4}_{ab}(\tau,\tau')G^{\circ 4}_{ab}(\tau,\tau') \notag\\
 &-\int d\tau\sum_{ia} s_{ia}\big(-\frac{M}{2}\partial^2_{\tau} + \frac{z}{2}\big)s_{ia}(\tau)+ \frac{NJ^2}{4}\int d\tau d\tau' \sum_{ab}Q^{\circ 3}_{ab}(\tau,\tau') 
 \bigg]\delta\bigg(NQ_{ab}-\sum_{i}s_{ia}s_{ib} \bigg)\delta\bigg(NG_{ab}-\sum_{i}\chi_{ia}\chi_{ib} \bigg)
\end{align}
To promote $G, Q$ as fluctuating dynamic fields, conjugate fields are introduced by using the relation $\delta\big(N G_{ab}(\t,\t')-\sum_{m}\chi_{ma}(\t)\chi_{mb}(\t')\big)=\int \mathcal{D}[\Sigma(\t,\t')] \exp[+\int d\t d\t'\Sigma_{ab}(\t,\t')\big(N G_{ab}(\t,\t')-\sum_{m}\chi_{ma}(\t)\chi_{mb}(\t')\big)]$ for $G_{ab}(\t,\t')$ and similarly for $Q_{ab}(\t,\t')$, $\delta\big(N Q_{ab}(\t,\t')-\sum_{i}s_{\gamma a}(\t)s_{i b}(\t')\big)=\int \mathcal{D}[\Pi(\t,\t')] \exp[-\int d\t d\t' (\Pi_{ab}(\t,\t')/2)\big(N Q_{ab}(\t,\t')-\sum_{i}s_{i a}(\t)s_{i b}(\t')\big)]$, where $\Pi(\t,\t')$ as the conjugate field. Next, we integrate over $s,\chi$ and get:
\begin{align}
    \overline{Z^r} &= \int \mathcal{D}[G, \Sigma, Q, \Pi] e^{-S_{\rm eff}[G, \Sigma, Q, \Pi]}
\end{align}
The effective action $S_{\rm eff}$ reads as:
\begin{align}
S_{\rm eff} &= -N \operatorname{Tr} 
             \ln \big( -G^{-1}_{0,ab} + \Sigma_{ab} \big) 
             - \frac{N}{2} \operatorname{Tr} \ln \big( Q^{-1}_{0,ab} -\Pi_{ab} \big) 
             + \frac{NV^2}{4}\sum_{ab}\int d\tau d\tau' G^{\circ 4}_{ab}(\tau,\tau') Q^{\circ 4}_{ab}(\tau,\tau') \notag \\
           &\quad -N \int d\tau d\tau' \sum_{ab}\Sigma_{ab}(\tau,\tau') G_{ab}(\tau,\tau')
             - \frac{N}{2} \int d\tau d\tau' \sum_{ab}\Pi_{ab}(\tau,\tau') Q_{ab}(\tau,\tau') 
             + \int d\tau d\tau' \frac{J^2}{4} \sum_{ab}Q^{\circ 3}_{ab}(\tau,\tau')
\end{align}
where 
\begin{align}
    G^{-1}_{0, ab} &= -\partial_{\tau}\delta(\tau-\tau')\delta_{ab} \\
    Q^{-1}_{0,ab} &= \big(-\frac{1}{\Gamma}\partial^2_{\tau}+z)\delta(\tau-\tau')\delta_{ab}
\end{align}
In the above derivation, we have set $\hbar=1$, and introduced a quantum parameter $\Gamma=1/MJ$. Taking the variation of $S_{\rm eff}$ with respect to $X(=G, Q, \Sigma, \Pi)$ and setting $\partial S_{\rm eff}/\partial X_{ab} = 0$, we get: 

\begin{align}
    Q^{-1}_{ab}(\tau,\tau') &= Q^{-1}_0(\tau,\tau') \delta_{ab}- \Pi_{ab}(\tau,\tau') \\
    \Pi_{ab}(\tau,\tau') &= \frac{3J^2}{2}Q^2_{ab}(\tau,\tau') + V^2 Q^3_{ab}G^4_{ab}(\tau,\tau')\\
    G^{-1}_{ab}(\tau,\tau') &= G^{-1}_0(\tau,\tau')\delta_{ab}-\Sigma_{ab} (\tau,\tau') \\
    \Sigma_{ab}(\tau,\tau') &= V^2 Q^{4}_{ab}(\tau,\tau') G^{3}_{ab}(\tau,\tau') 
\end{align}

\subsection{Coupled Model: Paramagnetic phase}
As discussed in the main text, the fermionic Green's function always takes the diagonal form: 
\begin{align}
    G_{ab}(\tau) &= G(\tau) \delta_{ab}
\end{align}
and the bosonic Green's function ($Q$) has diagonal form in paramagnetic phase:  
\begin{align}
    Q_{ab}(\tau) &= Q(\tau) \delta_{ab}
\end{align}
In this case, the saddle point equations become:
\begin{align}
    Q^{-1}(\tau) &= Q^{-1}_{0}(\tau)-\Pi(\tau) \label{eqn:dyson_Qdiag}\\
    \Pi(\tau) &= \frac{3J^2}{2}Q^2(\tau)+V^2Q^3(\tau)G^4(\tau)\label{eqn:selfE_Pidiag}\\
     G^{-1}(\tau) &= G^{-1}_0(\tau) - \Sigma(\tau) \\
    \Sigma(\tau) &= V^2 Q^{4}(\tau) G^{3}(\tau)   
\end{align}

\subsection{Coupled Model: Spin-glass phase}
As before, the fermionic Green's function takes the diagonal form: 
\begin{align}
    G_{ab}(\tau) &= G(\tau) \delta_{ab}
\end{align}
and the bosonic Green's function ($Q$) takes the 1-RSB form:  
\begin{align}
    Q_{ab}(\tau) &= (Q_d(\tau)-q_{\rm EA})\delta_{ab}+q_{\rm EA}\epsilon_{ab}
\end{align}
where $\epsilon_{ab}=1$ for $a, b$ in diagonal block, otherwise it is zero. Here, $Q_d(\tau)=Q_{aa}(\tau)$ is diagonal component. In this case, the saddle-point equations become:
\begin{align}
    Q^{-1}_{ab}(\tau) &= Q^{-1}_{0}(\tau)\delta_{ab}-\Pi_{ab}(\tau) \label{eqn:dyson_Q}\\
    \Pi_{ab}(\tau) &= \frac{3J^2}{2}Q^2_{ab}(\tau)+V^2Q^3_{ab}(\tau)G^4(\tau)\delta_{ab}\label{eqn:selfE_Pi}\\
     G^{-1}(\tau) &= G^{-1}_0(\tau) - \Sigma(\tau) \\
    \Sigma(\tau) &= V^2 Q^{4}_{aa}(\tau) G^{3}(\tau)   
\end{align}

To write down explicitly the saddle point equations for $Q_{ab}$ for $a=b$ and $a\neq b$, we write  the above 1-RSB order parameter in Matsubara frequency: 
	\beq \label{rsb1}
	\tilde{Q}_{ab}(\w_k) = (\tilde{Q}_d(\w_k)-\qt)\delta_{ab}+\qt\epsilon_{ab},
	\eeq
	where $\qt=\beta \qe \delta_{\w_k, 0}$ and $\tilde{Q}_d(\w_k)$ is Matsubara Fourier transformation of $Q_d(\tau)=Q_{aa}(\tau)$. Now, it is convenient to write $Q_d(\tau) = \qr(\tau) + \qe$.
The inverse matrix ${\tilde{Q}^{-1}}(\w_k)$ has the following structure~\cite{Leticia2001}
\beq \label{rsb2}
{\ginv}(\w_k) = A(\w_k)\delta_{ab} + B(\w_k)\epsilon_{ab}
\eeq 
with
\beq \label{rsb3}
A(\w_k) = \frac{1}{\tilde{\qd}(\w_k)-\qt}, \hspace{0.8cm}
B(\w_k) = \frac{-\qt}{\tilde{\qd}(\w_k)^2-(m-1)\qt^2+(m-2)\tilde{\qd}(\w_k)\qt}.
\eeq  
Here, $m$ is the break point and the above result for $A(\w_k), B(\w_k)$ is obtained after taking the replica limit $r\to 0$.   
The diagonal element of the inverse matrix (${\tilde{Q}^{-1}}$) is given by
\beq \label{rsb4}
(\tilde{Q}^{-1})_{aa}(\w_k) = A(\w_k) + B(\w_k) = \frac{\tilde{\qd}(\w_k)+(m-2)\qt}{(\tilde{\qd}(\w_k)^2+(m-2)\qt\tilde{\qd}(\w_k)-(m-1)\qt^2)}.
\eeq      
The off-diagonal element ($a\neq b$) is 
\beq \label{rsb5}
\ginv(\w_k) = B(\w_k).
\eeq   
The saddle-point equation for diagonal component $\tilde{\qd}(\w_k)$ can be obtained by using equations \eqref{eqn:dyson_Q} and \eqref{rsb4} as 
	\beq \label{rsb6}
	M{\w_k^2}+z = \frac{\tilde{\qd}(\w_k)+(m-2)\qt}{(\tilde{\qd}(\w_k)^2+(m-2)\qt\tilde{\qd}(\w_k)-(m-1)\qt^2)} + \Pi(\w_k),
	\eeq 
with $\Pi(\w_k)=\Pi_{aa}(\w_k)$ is the Fourier transform of $\Pi_{aa}(\tau)$ which reads as:
\begin{align}
    \Pi_{aa}(\tau) &= \frac{3J^2}{2}Q^2_{d}(\tau)+V^2Q^3_{d}(\tau)G^4(\tau) 
\end{align}
The saddle point equation for $\qe$ is obtained using eqn.\ref{rsb5} and eqn.\ref{eqn:selfE_Pi}
{\beq \label{rsb7}
	0 =-\frac{\qe}{\tilde{\qd}(0)^2-(m-1)\beta^2\qe^2+(m-2)\beta\tilde{\qd}(0)\qe} + \frac{ p}{2} \qe^{p-1}.
	\eeq }

Using eqn. \ref{rsb7}, we can write down the diagonal saddle point equation more conveniently with ${\qr}(\w_k)$. This reads as:
\begin{align}
    M\w^2_k + z &= \frac{1}{\qr(\w_k)} + \Pi_{\rm reg}(\w_k) 
\end{align}
where 
\begin{align}
    \Pi_{\rm reg}(\tau) &= \frac{3J^2}{2}\qd^2(\tau) + V^2 \qd^3(\tau) G^4(\tau) - \frac{3J^2}{2}\qe^2 
\end{align}
\subsubsection*{Thermodynamic spin-glass phase and marginal spin-glass phase }
In the thermodynamic spin-glass phase, the breakpoint parameter $m$ is treated as a variational parameter that minimizes the free-energy functional.  
The value of $m$ is determined by explicitly considering the $m$-dependence of the free-energy functional $f[m]$.  
The boson–fermion coupling term 
\[
\frac{NV^2}{4}\sum_{ab} \int d\tau d\tau' \, G^{\circ 4}_{ab}(\tau,\tau') \, Q^{\circ 4}_{ab}(\tau,\tau')
\propto r \cdot \frac{NV^2}{4} \int d\tau d\tau' \, G^{\circ 4}(\tau,\tau') \, Q^{\circ 4}_{d}(\tau,\tau')
\]
reduces to a term proportional to the replica index $r$ because $G_{ab}(\tau) = G(\tau)\delta_{ab}$ is diagonal in replica space.  
As a result, this contribution doesnot depend  on $m$, and the variation of $f[m]$ reproduces the same set of equations as in the decoupled ($V=0$) $p$-spin-glass model.  
We refer the reader to Ref.~\cite{Leticia2001} for a detailed derivation of these equations.  
For $p=3$, the equation for $m$ relating it to the Edwards–Anderson parameter $q_{\rm EA}$ is given by
\begin{align}\label{eq:qe-m}
    \frac{p(\beta m)^2}{2}\, q_{\rm EA}^p &= \frac{2x_p^2}{1+x_p},
\end{align}
where $x_p=\frac{my}{1-y}$, $y=\frac{\beta\qe}{\tilde{Q}_d(0)}$, is determined by
\begin{align}
   \ln \bigg(\frac{1}{1+x_p}\bigg) + \frac{x_p}{1+x_p} + \frac{x_p^2}{p(1+x_p)} &= 0.
\end{align}
For $p=3$, it yields $x_p = 1.81696$.

In the marginal spin-glass phase, the breakpoint parameter $m$ is determined by imposing the marginality condition, which requires the replicon eigenvalue $\Lambda_T = 0$.  
To compute the replicon eigenvalue, one considers the second-order variation of the free-energy functional with respect to $Q_{ab}$ for $a \neq b$.  
For the same reason discussed in the thermodynamic spin-glass case, the boson–fermion coupling term does not modify the replicon eigenvalue which is given by\cite{Leticia2001}:
\begin{align}
    \Lambda_T &= \beta^2 \bigg[\frac{1}{\tilde{Q}_d(0)-\qt } - \frac{p(p-1)}{2}\qe^{p-2} \bigg]
\end{align}
Consequently, the marginality condition remains identical to that of the decoupled case.  
In the marginal spin-glass phase, by setting $\Lambda_T=0$, one obtain $x_p = p - 2$, and the relation between $m$ and $q_{\rm EA}$ is the same as given in Eq.~\ref{eq:qe-m}.

\section{Perturbative correction to fermionic Green's function in quantum paramagnetic regime}
\label{app:pert_self_energy_qpm}

In this appendix we present a simple first order perturbative estimate of the leading  fermionic self-energy correction in the quantum paramagnetic regime, where the bosonic correlator is short-ranged in imaginary time due to a gap. We work in the low-energy limit $\beta\to\infty$ so that imaginary time becomes continuous.

We assume that in the quantum paramagnet the bosonic Green's function decays exponentially,
\begin{equation}
Q(\tau)\simeq e^{-\Delta|\tau|}, \qquad \Delta>0,
\label{eq:Q_exp}
\end{equation}
which captures the presence of a bosonic gap $\Delta$ and ensures that the bosonic correlations are short-ranged in $\tau$. For the fermions we use the bare (non-interacting) Green's function at $T=0$,
\begin{equation}
G_0(\tau)=\frac{1}{2}\,\mathrm{sgn}(\tau),
\label{eq:G0_sign}
\end{equation}
so that in frequency space $G_0(i\omega)=1/(-\ci\omega)$.
The leading self-energy due to the boson--fermion coupling $V$ takes the schematic form
\begin{equation}
\Sigma(\tau)=V^2\,Q^4(\tau)\,G^3(\tau).
\label{eq:Sigma_tau_def}
\end{equation}
As a first order perturbative estimate, we evaluate $\Sigma$ by inserting $Q(\tau)$ from Eq.~\eqref{eq:Q_exp} and $G(\tau)\approx G_0(\tau)$ from Eq.~\eqref{eq:G0_sign} into Eq.~\eqref{eq:Sigma_tau_def}.

Using $Q^4(\tau)=e^{-4\Delta|\tau|}$ and $G_0^3(\tau)=(1/8)\,\mathrm{sgn}(\tau)$, we obtain
\begin{equation}
\Sigma^{(1)}(\tau)
=\frac{V^2}{8}\,e^{-4\Delta|\tau|}\,\mathrm{sgn}(\tau).
\label{eq:Sigma_tau_result}
\end{equation}
Fourier transforming Eq.~\eqref{eq:Sigma_tau_result} to continuous Matsubara frequency,
\begin{equation}
\Sigma^{(1)}(i\omega)=\int_{-\infty}^{\infty} d\tau\, e^{i\omega\tau}\,\Sigma^{(1)}(\tau),
\end{equation}
we use the standard integral (for $a>0$)
\begin{equation}
\int_{-\infty}^{\infty} d\tau\; e^{i\omega\tau}\,e^{-a|\tau|}\,\mathrm{sgn}(\tau)
=\frac{2i\omega}{a^2+\omega^2}.
\label{eq:FT_identity}
\end{equation}
Setting $a=4\Delta$ yields
\begin{equation}
\Sigma^{(1)}(i\omega)
=\frac{V^2}{8}\,\frac{2i\omega}{(4\Delta)^2+\omega^2}
=\frac{V^2}{4}\,\frac{i\omega}{\omega^2+16\Delta^2}.
\label{eq:Sigma_iw_result}
\end{equation}
Thus, the leading perturbative self-energy is analytic at low frequency and is proportional to $i\omega$. The fermionic Dyson (Schwinger--Dyson) equation reads
\begin{equation}
G(i\omega)=\frac{1}{-\ci\omega-\Sigma(\ci\omega)}.
\label{eq:Dyson}
\end{equation}
Inserting Eq.~\eqref{eq:Sigma_iw_result} and expanding to leading order in $V^2$ gives
\begin{align}
G^{(1)}(i\omega)
&=
\frac{1}{-\ci\omega}\,
\frac{1}{1+\dfrac{V^2}{4(\omega^2+16\Delta^2)}} \nonumber\\
&\simeq
\frac{1}{-\ci\omega}\left[1-\frac{V^2}{4(\omega^2+16\Delta^2)}+O(V^4)\right].
\label{eq:G_iw_pert}
\end{align}
In particular, in the low-frequency regime $|\omega|\ll \Delta$, we find
\begin{equation}
G^{(1)}(i\omega) 
\simeq \frac{1}{-i\omega}\left(1-\frac{V^2}{64\Delta^2}+O(V^4)\right)
\label{eq:G_loww_pert}
\end{equation}

Equation~\eqref{eq:Sigma_iw_result} shows that for an exponentially decaying bosonic correlator the leading fermionic self-energy is regular (analytic) at $\omega\to 0$, producing a conventional renormalization of the $i\omega$ term rather than any non-analytic marginal scaling. Therefore, the imaginary time Green's function reads at $T\to 0$ limit as below: 

\begin{align}
    G^{(1)}(\tau) \simeq \frac{1}{2}\big(1- \frac{V^2}{64\Delta^2}\big) \mathrm{sgn}(\tau)
\end{align}

\section{The SYK Green's function for $\mathcal{J}\to 0$ limit}\label{sec:G0-tau}
\begin{figure}[H]
    \centering
    \includegraphics[width=0.65\linewidth]{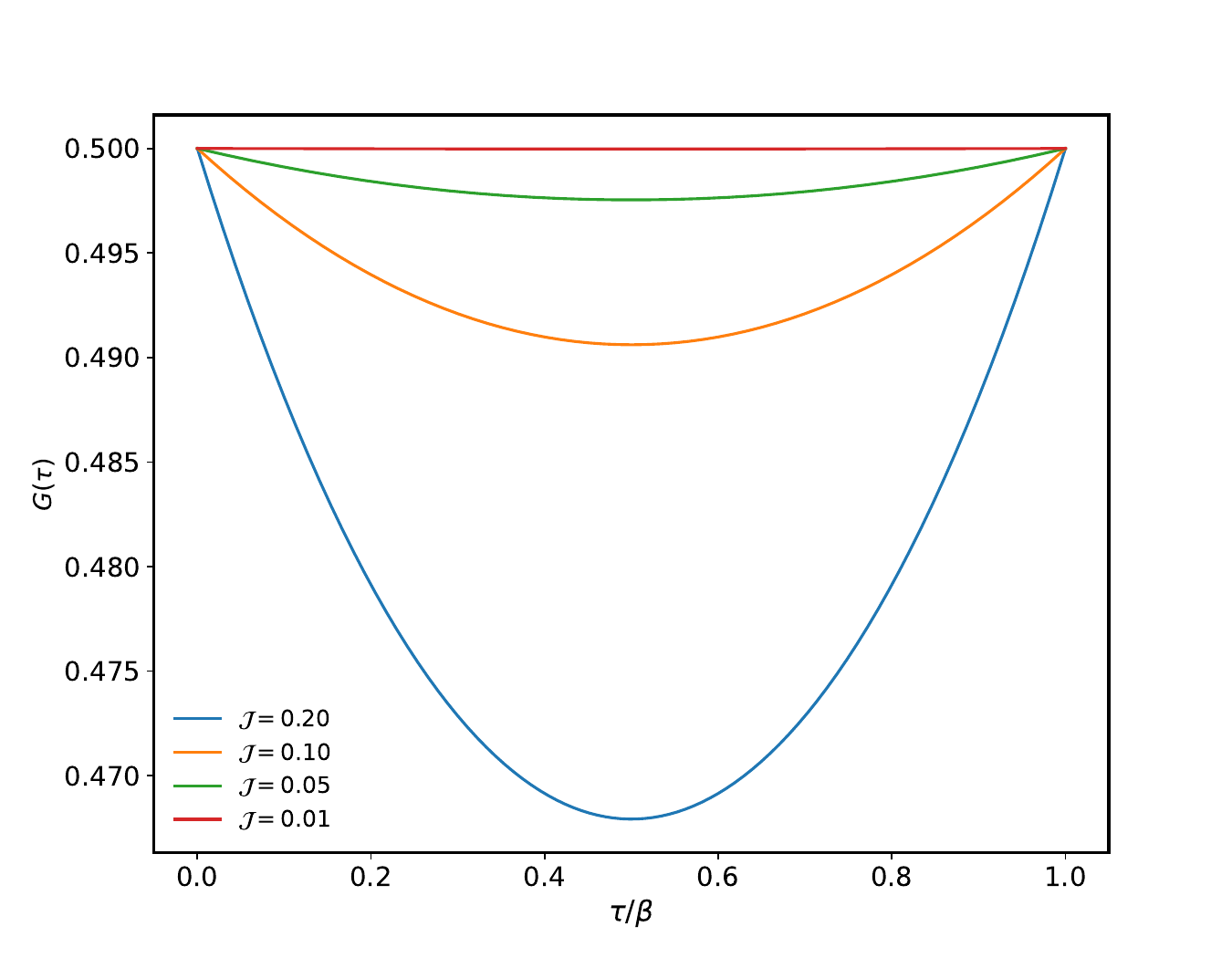}
    \caption{The pure SYK Green's function $G(\tau)$ for coupling strength $\mathcal{J}=0.2, 0.1, 0.05, 0.01$ are shown for $T=0.125$}
    \label{fig:G0-tau}
\end{figure}

In Fig.~\ref{fig:G0-tau}, we present the pure SYK Green's function $G_0(\tau)$ as a function of $\tau$ for $T=0.125$ and several small coupling strengths $\mathcal{J} = 0.2, 0.1, 0.05, 0.01$. As the coupling strength $\mathcal{J} \to 0$, $G_0(\tau)$ smoothly approaches the free-fermion Green's function, $G(\tau) = \frac{1}{2}\,\text{sign}(\tau)$, as expected.

\section{Quantitative characterization of the crossover scale separating SYK and non-SYK behavior
}\label{Appendix:CrossoverT}

In this appendix, we provide a quantitative characterization of the crossover temperature scale $T^*(V,\Gamma)$ separating the low-temperature SYK-like regime from the intermediate-temperature regime with strong deviations from the SYK behavior in the spin-glass phase as we discussed in the main text.

Our strategy is to compare the imaginary-time fermionic Green's function $G(\tau)$ obtained from the full interacting model with a reference SYK Green's function $G_0(\tau)$, computed for a pure SYK model with an effective coupling $J_{\rm eff}=V\cdot \langle Q^2(\tau)\rangle $ chosen to match the low-temperature scale of the interacting theory. The crossover temperature $T^*$ is then roughly defined as the temperature above which the shape of $G(\tau)$ departs appreciably from the SYK form.

\subsection{Normalization of Green's functions}

Since the comparison focuses on the \emph{functional form} of the Green's functions rather than their absolute difference, we first rescale both $G(\tau)$ and $G_0(\tau)$ using a min--max normalization,
\begin{equation}
\tilde{f}(\tau)
=
\frac{f(\tau)-f_{\min}}{f_{\max}-f_{\min}},
\label{eq:minmax}
\end{equation}
where $f(\tau)$ denotes either $G(\tau)$ or $G_0(\tau)$, and $f_{\min}$ and $f_{\max}$ are the minimum and maximum values of $f(\tau)$ over $\tau\in[0,\beta]$. This procedure ensures that both rescaled Green's functions lie in the interval $[0,1]$, allowing for a direct comparison of their shapes independent of absolute amplitude differences for different $T, \Gamma$.

\subsection{Total deviation measure}

To quantify the deviation between the two rescaled Green's functions, we introduce a total variation distance (TVD), defined as
\begin{equation}
\mathrm{TVD}
= 
\int_0^1 d  \tilde{\tau}
\left|
\tilde{G}(\tilde{\tau})-\tilde{G}_0(\tilde{\tau})
\right|.
\label{eq:tdv}
\end{equation}
This measure captures the integrated pointwise difference between the two curves over the full imaginary-time interval  and provides a simple scalar quantity characterizing their similarity. Here, $\tilde{\tau}=\tau/\beta$. 

We interpret $\mathrm{TVD}$ as a distance in function space: small values indicate that $G(\tau)$ closely follows the SYK form, while larger values signal a significant departure from SYK behavior.

\subsection{Definition of the crossover temperature}

\begin{figure}
    \centering
    \includegraphics[width=0.85\linewidth]{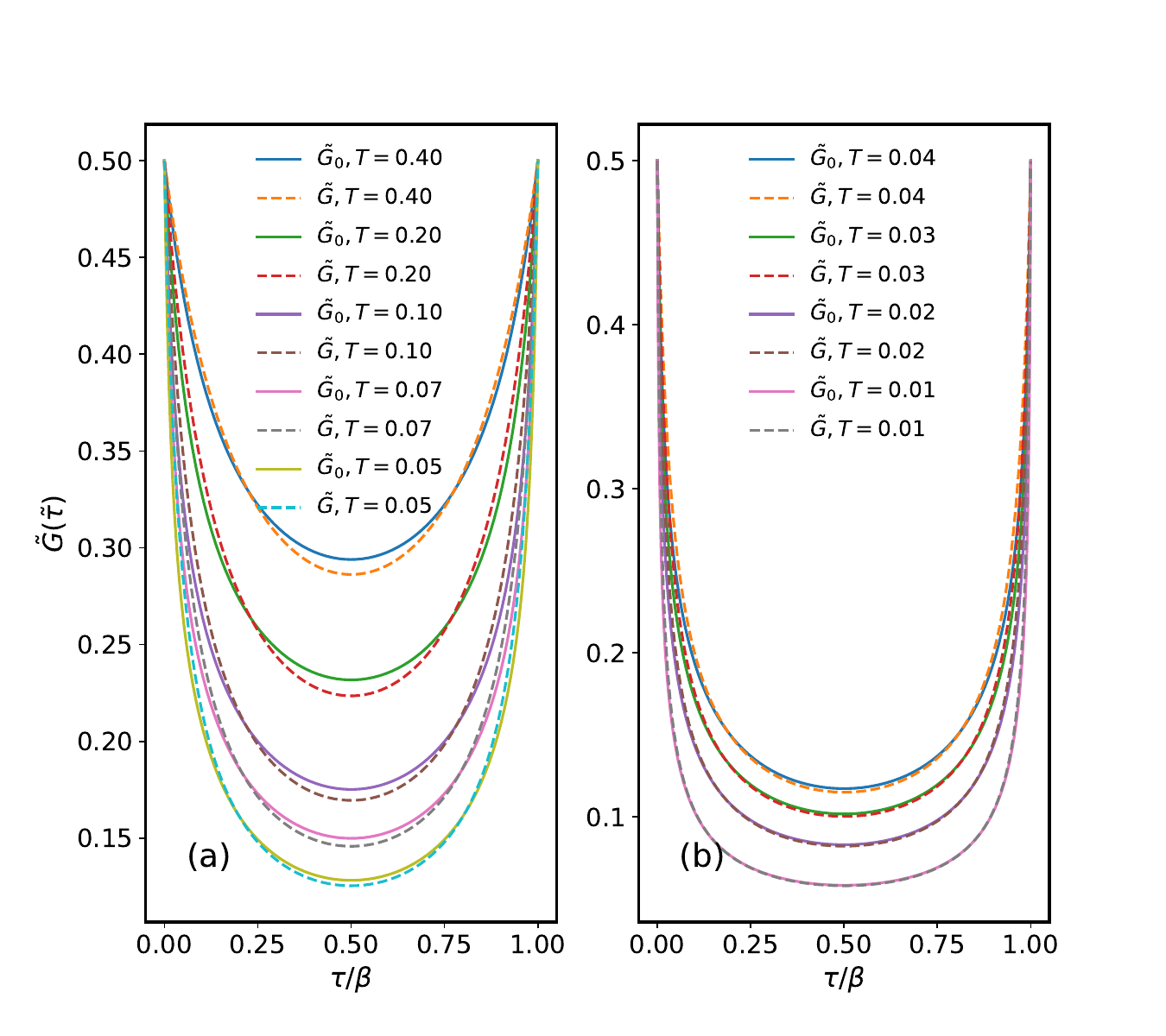}
    \caption{The scaled fermionic Green’s function $\tilde{G}(\tilde{\tau})$ (solid line) for the coupled model is compared with the scaled SYK Green’s function $\tilde{G}_0(\tilde{\tau})$ (dotted line) evaluated using a static effective coupling $\mathcal{J}_{\rm eff}=V\langle Q_d^2(\tau)\rangle$. The left panel corresponds to a relatively higher temperature, while the right panel shows results at a lower temperature, for $\Gamma=0.8$ and $V/J=5.0$. The deviation of $\tilde{G}$ from the SYK Green’s function $\tilde{G}_0$ begins approximately around $T \approx 0.05$.
 }
    \label{fig:ScaledG_gamma0p8}
\end{figure}

\begin{figure}[htb]
    \centering
    \includegraphics[width=0.85\linewidth]{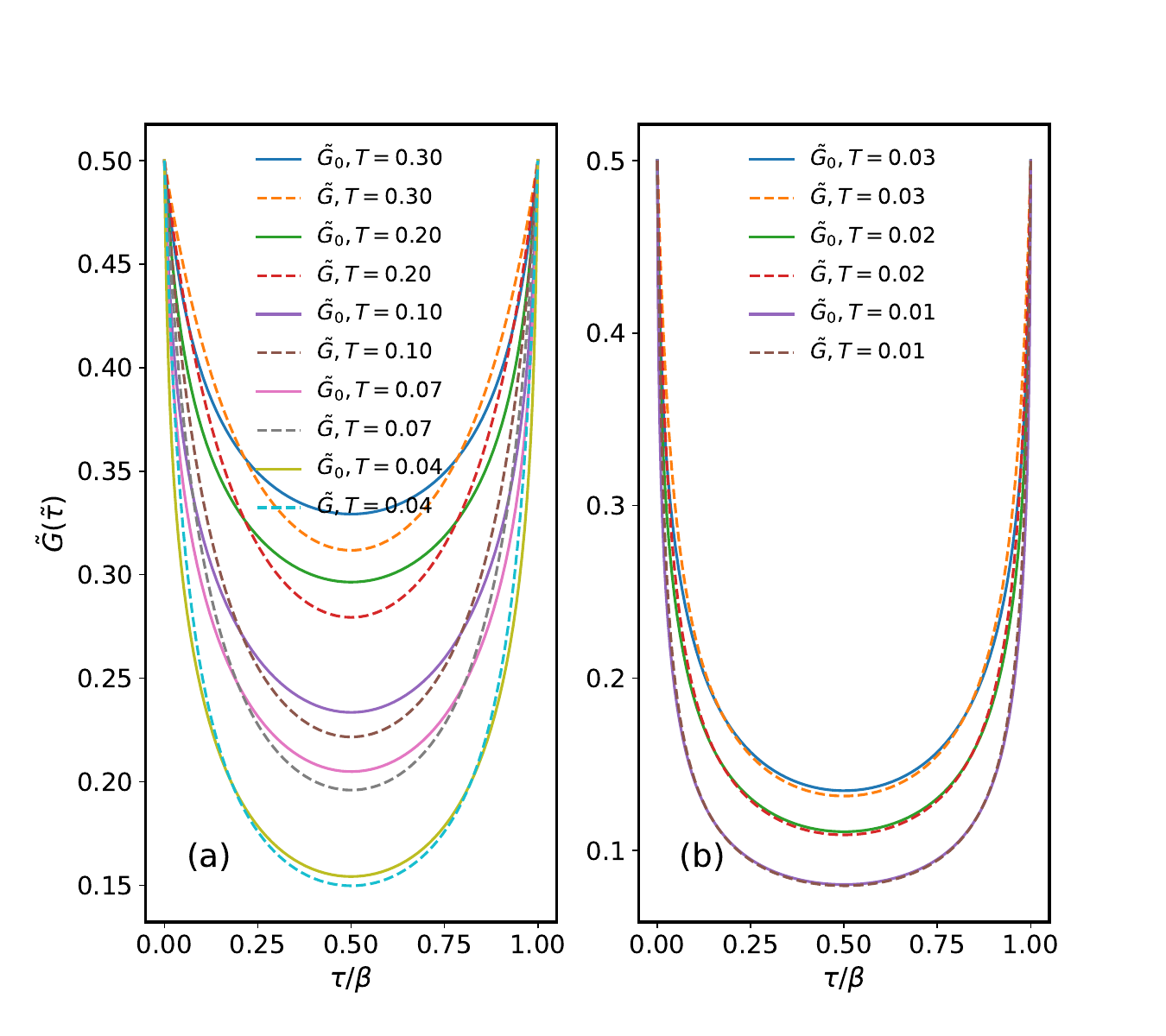}
    \caption{The scaled fermionic Green’s function $\tilde{G}(\tilde{\tau})$ (solid line) for the coupled model is compared with the scaled SYK Green’s function $\tilde{G}_0(\tilde{\tau})$ (dotted line) evaluated using a static effective coupling $\mathcal{J}_{\rm eff}=V\langle Q_d^2(\tau)\rangle$. The left panel corresponds to a relatively higher temperature, while the right panel shows results at a lower temperature, for $\Gamma=2.0$ and $V/J=5.0$. The deviation of $\tilde{G}$ from the SYK Green’s function $\tilde{G}_0$ begins at approximately $T \approx 0.03$
 }
    \label{fig:ScaledG_gamma2p0}
\end{figure}

We define the crossover temperature $T^*(V,\Gamma)$ as the temperature at which the total variation distance (TVD) exceeds a fixed tolerance $\epsilon$,
\begin{equation}
\mathrm{TVD} > \epsilon.
\end{equation}
The threshold $\epsilon$ is chosen empirically to distinguish regimes in which the interacting Green’s function and the SYK reference Green’s function are visually and quantitatively indistinguishable from those where clear deviations are present.

To determine an appropriate tolerance, we compare the scaled Green’s function $\tilde{G}(\tilde{\tau})$ with the scaled SYK Green’s function $\tilde{G}_0(\tilde{\tau})$ and identify the temperature at which noticeable deviations first appear. From this analysis, we find that the corresponding TVD is approximately $\epsilon \simeq 0.01$ across the explored parameter ranges of $T$, $\Gamma$, and $V$.
 We have verified that modest variations of $\epsilon$ do not qualitatively affect the extracted crossover trends.

For temperatures $T<T^*$, the Green's function $G(\tau)$ agrees well with the SYK Green's function $G_0(\tau)$, indicating SYK-like fermionic behavior. For $T>T^*$, the deviation increases rapidly, signaling a strong deviation from SYK universality due to the coupling to the bosonic sector. Representative examples are shown in Figs.~\ref{fig:ScaledG_gamma0p8} and \ref{fig:ScaledG_gamma2p0}, where we plot the scaled Green’s functions for $\Gamma=0.8$ and $\Gamma=2.0$ at $V/J=5.0$. In these cases, the visually identified deviation temperature corresponds to $\mathrm{TVD} \approx 0.01$.

\subsection{Dependence of $T^*$ on model parameters}

\begin{figure}[htb]
    \centering
    \includegraphics[width=0.85\linewidth]{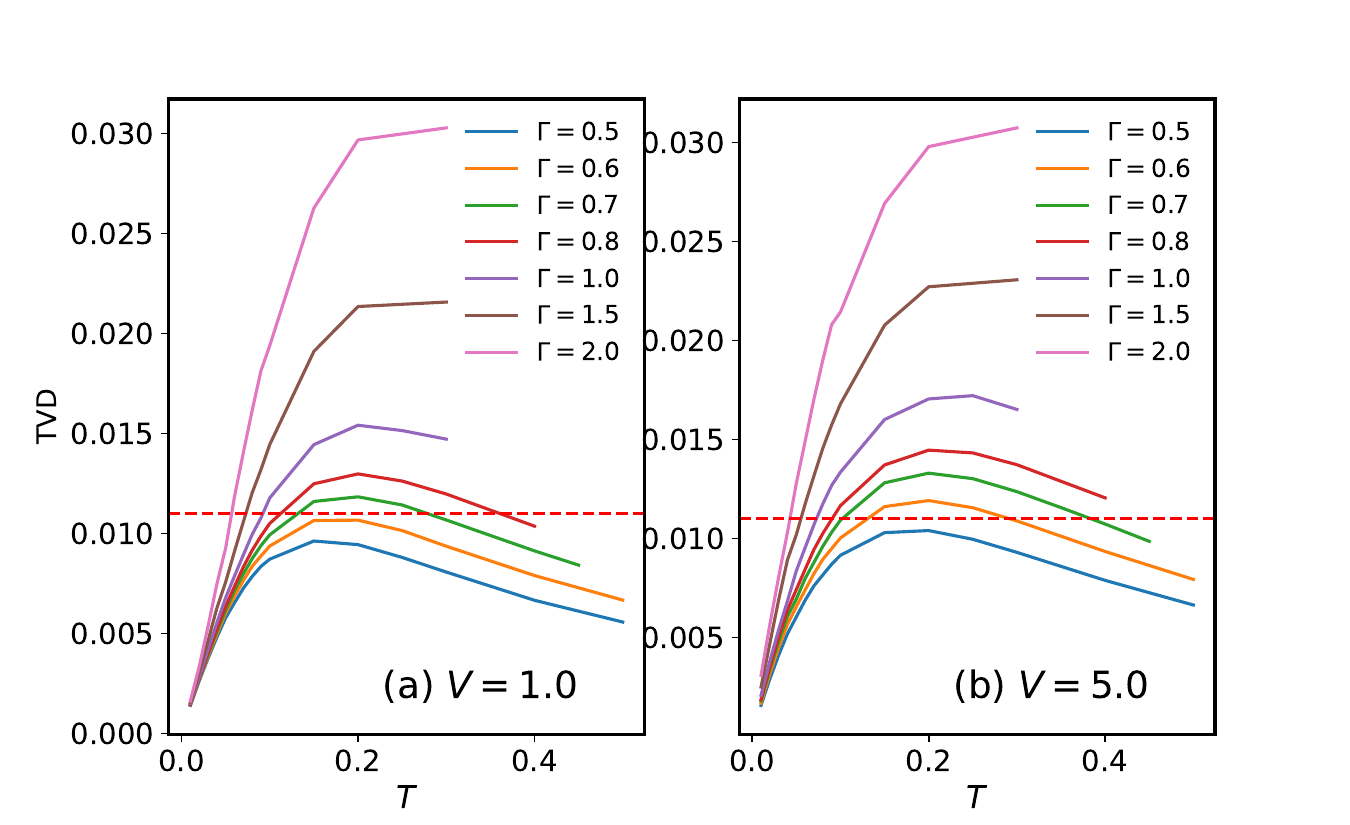}
    \caption{The total variation distance (TVD), defined as the absolute difference between $G(\tau)$ and $G_0(\tau)$ computed with a static effective coupling, is plotted as a function of temperature $T$ for different values of $\Gamma$. Results are shown for two interaction strengths: $V=1$ in panel (a) and $V=5$ in panel (b).
 }
    \label{fig:TVD_vs_T}
\end{figure}

Applying this procedure across a range of coupling strengths $V$ and quantum fluctuation parameters $\Gamma$, we extract the crossover scale $T^*(V,\Gamma)$. We find that $T^*$ depends only weakly on $V$ over the explored parameter range, while exhibiting a pronounced dependence on $\Gamma$. In particular, increasing $\Gamma$ suppresses $T^*$, thereby shifting the onset of deviations from SYK behavior to progressively lower temperatures. These results are summarized in Fig.~\ref{fig:TVD_vs_T}, where we plot the total variation distance (TVD) as a function of temperature for representative values of $V$ and $\Gamma$. The extracted crossover scales $T^*(\Gamma)$ for different values of $V$ are shown in Fig.~\ref{fig:crossoverT}.

For $\Gamma \lesssim 0.5$, the TVD remains below the threshold $\epsilon$ over the entire accessible temperature range, indicating that $G(\tau)$ retains an SYK-like form. This behavior is consistent with our intuitive expectations in the classical limit $\Gamma \to 0$, as discussed in the main text. For intermediate values, $\Gamma=0.6$ and $0.7$, we observe a more intricate temperature dependence: at very low temperatures, $G(\tau)$ closely matches the SYK reference Green’s function $G_0(\tau)$; upon increasing temperature, a crossover occurs in which $G(\tau)$ deviates from SYK behavior; and at sufficiently high temperatures, $G(\tau)$ again approaches $G_0(\tau)$. This high-temperature recovery is expected, as thermal fluctuation make the bosonic correlator $Q(\tau)$ slowly varying, thereby reducing its influence on the fermionic dynamics.

Such high-$T$ SYK-like behavior is not expected to persist at larger values of $\Gamma$. Indeed, for sufficiently large $\Gamma$, the spin-glass (SG) phase is confined to a narrow low-temperature window and does not extend to high temperatures.

Overall, this analysis provides a quantitative characterization of the crossover temperature that separates the low-temperature SYK regime from the intermediate-temperature non-SYK regime.

\section{Scaling of fermion Green's function in the spin glass phase}\label{sec:scaling-SYK-tSG}
The pure SYK Green's function admits a conformal solution at low energies, decaying as a power law:
\begin{align}
    G_c(\tau) &= \frac{b}{|\mathcal{J} \tau|^{2\Delta}} \, \text{sign}(\tau),
\end{align}
where $\Delta = 1/4$ for the four-fermion SYK model given in Eq.~\ref{eq:SYKHam}. Using a conformal mapping, the finite-temperature solution takes the form
\begin{align}
    G_c(\tau) &= b \bigg[\frac{\pi}{\sin\!\big(\pi \tau/\beta\big)}\bigg]^{2\Delta} \text{sign}(\tau).
\end{align}

Numerically, the power-law behavior can be verified by plotting $1/G(\tau)$ as a function of $\sin(\pi \tau/\beta)$ on a log–log scale. For different temperatures $T$ or couplings $\mathcal{J}$, the pure SYK Green's function exhibits a linear region in the log–log plot at large times ($\tau \sim \beta/2$) with an identical slope, reflecting the universal power $2\Delta$. As $T$ increases, the range over which this linear behavior is observed becomes progressively narrower.

\begin{figure}[htb]
    \centering
    \includegraphics[width=0.85\linewidth]{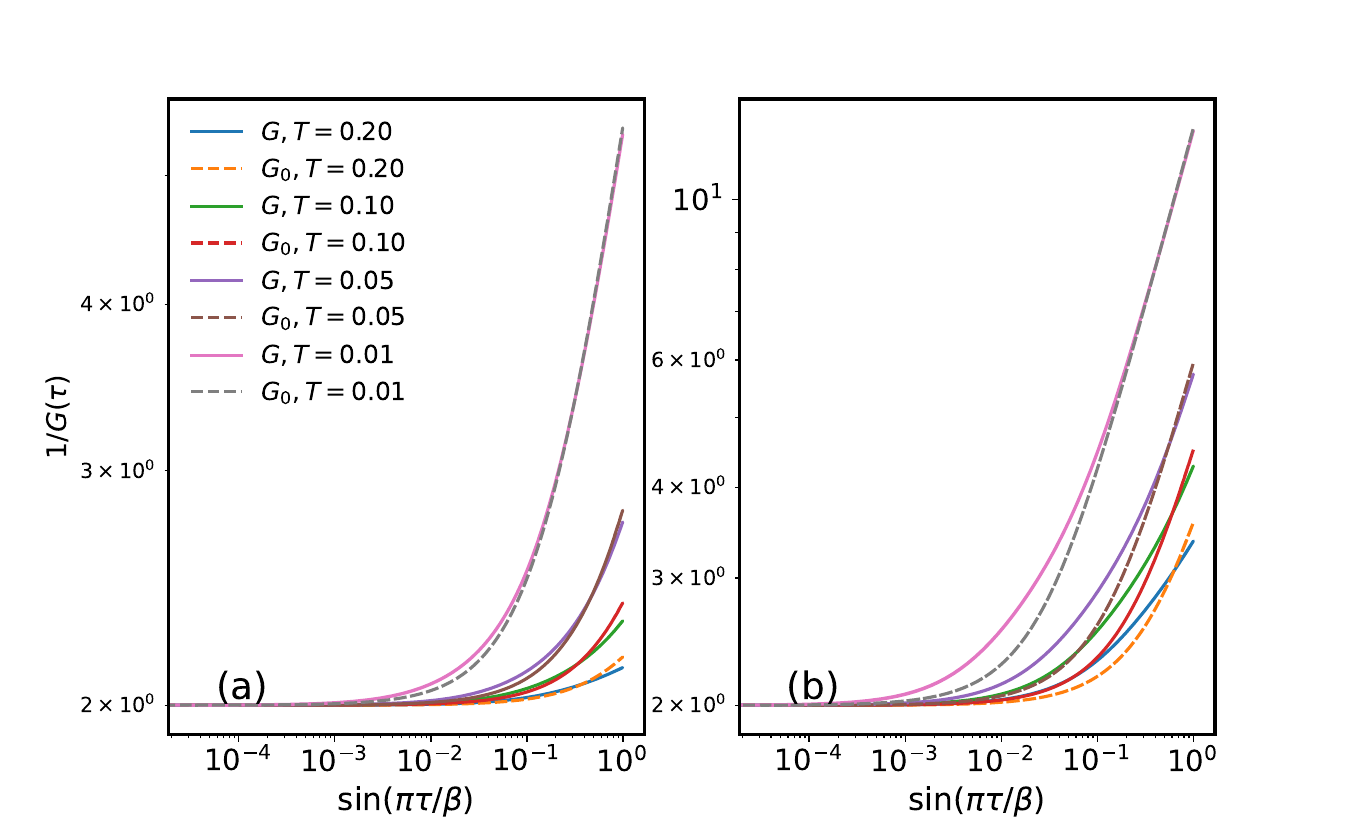}
    \caption{The fermionic Green's function $G(\tau)$ is shown in Fig.~\ref{fig:Q-mSG-J2p0}(a) and (b), where $1/G(\tau)$ is plotted as a function of $\sin(\pi \tau / \beta)$ on a log–log scale. Panels (a) and (b) correspond to coupling strengths $V/J = 1.0$ and $V/J = 5.0$, respectively, at fixed $\Gamma = 2.0$. For comparison, the Green's function of the pure SYK model with a static effective coupling $\mathcal{J}_{\rm eff}$ is also plotted.
}
    \label{fig:scaling-tSG-Gtau}
\end{figure}

In Fig.~\ref{fig:scaling-tSG-Gtau}, we present both the full fermionic Green's function $G(\tau)$ of the coupled problem and the pure SYK Green's function with a static renormalized coupling $\mathcal{J}_{\rm eff}=V\cdot \langle Q^2_d(\tau)\rangle$. Panels (a) and (b) correspond to $V/J=1.0$ and $V/J=5.0$, respectively, for $\Gamma=2.0$. At very low temperatures, $G(\tau)$ follows the conformal SYK scaling almost exactly at large $\tau$. However, at intermediate temperatures, the slope of $G(\tau)$ at large $\tau$ deviates from that of the SYK solution with static $\mathcal{J}_{\rm eff}$, signaling a departure from pure SYK behavior. As the linear scaling region becomes progressively narrower at higher $T$, we do not attempt a quantitative extraction of the power law exponent $2\Delta$.

\section{Wavefunctions}

 Although we shall not make  a detailed analysis of this here, we may guess what the wavefunctions look like at very low temperatures, and not too high values of $\Gamma$. 
Consistently with the discussion above, an approximate set of eigenvalues of the pure boson term is constructed with the lowest eigentates $|\alpha; a\rangle$ concentrated around each minimum $s^a$, where
$\alpha$ are quantum numbers. The lowest fermionic eigenstates within this state correspond to an effective  SYK model with effective couplings $\mathcal{J}^{(a)}_{mnpq}$. The set of eigenstates living  
in the metastable state $a$ are then constructed as combinations of tensor products of these two sets.
Eigentsates corresponding to different valleys are orthogonal both because their bosonic functions are, but also because the fermionic effective interactions $\mathcal{J}^{(a)}_{mnpq}$ that gave rise to them are very different. A better analysis may be done either by considering the real-time dynamics starting in various places, or the `Franz-Parisi construction', which still needs to be developed for quantum systems.

\clearpage
\twocolumngrid
\bibliography{ref}
\end{document}